\documentclass{aa}  
\usepackage{graphicx}
\usepackage{txfonts}
\usepackage{todonotes}
\usepackage[colorlinks=true, allcolors=blue]{hyperref}
\usepackage[nolist]{acronym}
\usepackage{comment}
\usepackage{booktabs}
\usepackage{longtable}
\usepackage{multirow}
\usepackage[switch]{lineno}
\usepackage{amssymb}
\usepackage{pifont}
\usepackage{caption}
\usepackage{subcaption}
\usepackage{wasysym}
\newcommand{\cmark}{\ding{51}}

\renewcommand{\epsilon}{\varepsilon}
\def\xmm{\textit{XMM-Newton}}
\def\chandra{\textit{Chandra}}
\def\swift{\textit{Swift}}

\def\fpeak{\textsc{Fpeak}}
\def\lpeak{\textsc{Lpeak}}

\def\EPVFLUX{4.5 -- 12 keV}
\def\EPVIIIFLUX{0.2 -- 12 keV}

\def\fluxunit{erg~cm$^{-2}$~s$^{-1}$}
\def\lumunit{erg~s$^{-1}$}

\def\logflux{log(flux)}
\newcommand{\solarM}{\,\mathrm{M}_\odot}

\DeclareMathOperator\erf{erf}

\title{The hunt for new pulsating ultraluminous X-ray sources: a clustering approach}
\titlerunning{The hunt for new PULXs: a clustering approach}

\authorrunning{N.\,O.\,Pinciroli Vago et al.}

\begin{document} 
   \author{N.\,O.\,Pinciroli Vago\inst{1,2}\thanks{Equal contribution.} \and
   R.\,Amato\inst{2}$^\star$ \and
   M.\,Imbrogno\inst{3,2} \and
   G.\,L.\,Israel\inst{2} \and 
   A.\,Belfiore\inst{4} \and
   K.\,Kovlakas\inst{5,6} \and 
   P.\,Fraternali\inst{1} \and
   M.\,Pasquato\inst{4,7}
   }

   \institute{Department of Electronics, Information and Bioengineering,
Politecnico di Milano, via G. Ponzio, 34, I-20133 Milan, Italy \\
   e-mail: \href{mailto:nicolooreste.pinciroli@polimi.it}{nicolooreste.pinciroli@polimi.it}
   \and  INAF--Osservatorio Astronomico di Roma, via Frascati 33, I-00078 Monte Porzio Catone, Italy 
    \and Dipartimento di Fisica, Università degli Studi di Roma ``Tor Vergata'', via della Ricerca Scientifica 1, I-00133 Roma, Italy
    \and INAF, Istituto di Astrofisica Spaziale e Fisica Cosmica, Via Alfonso Corti 12, I-20133 Milan, Italy 
    \and Institut d'Estudis Espacials de Catalunya (IEEC), Edifici RDIT, Campus UPC, 08860 Castelldefels (Barcelona), Spain
    \and Institute of Space Sciences (ICE, CSIC), Campus UAB, Carrer de Magrans, 08193 Barcelona, Spain
    \and Ciela – Montreal Institute for Astrophysical Data Analysis and Machine Learning, Montréal, Canada
}

  \date{Received DD Month YYYY; accepted DD Month YYYY}
 
  \abstract
  {The discovery of fast and variable coherent signals in a handful of ultraluminous X-ray sources (ULXs) testifies to the presence of super-Eddington accreting neutron stars, and drastically changed the understanding of the ULX class.  
  Our capability of discovering pulsations in ULXs is limited, among others, by poor statistics. However, catalogues and archives of high-energy missions, such as \xmm, \chandra\ and \swift, contain information, often overlooked, which can be used to identify new candidate pulsating ULXs (PULXs).}
  {The goal of this research is to single out candidate PULXs among those ULXs which have not shown pulsations due to an unfavourable combination of factors (low statistics, low pulsed fraction, etc.).}
  {We applied an Artificial Intelligence approach to an updated database of ULXs detected by \xmm. The sample counts 640 sources for a total of $\sim$1800 observations, 95 of which are those of known PULXs. We first used an unsupervised clustering algorithm to sort out sources with similar characteristics into two clusters.
   {Then, the sample of known PULX observations has been used to set the separation threshold between the two clusters and to identify the one containing the new candidate PULXs.}}
   {We found that only a few criteria are needed to assign the membership of an observation to one of the two clusters. Moreover, the best result, in terms of the capability of assigning all the known PULXs in one of the two clusters, is obtained when the maximum observed flux for each source is included in the clustering algorithm. The cluster of new candidate PULXs counts 85 unique sources for 355 observations, with $\sim$85\% of these new candidates having multiple observations. A preliminary timing analysis found no new pulsations for these candidates.}
   {This work presents a sample of new candidate PULXs observed by \xmm, the properties of which are similar (in a multi-dimensional phase space) to those of the known PULXs, despite the absence of pulsations in their light curves. While this result is a clear example of the predictive power of non-traditional, AI-based methods, it also highlights the need for high-statistics observational data to reveal coherent signals from the sources in this sample and thus validate the robustness of the approach.}

   \keywords{Astronomical data bases -- Methods: statistical -- Catalogs -- X-rays: binaries}

   \maketitle

\acresetall

\section{Introduction}

\acp{ULX} are a class of extra-galactic, point-like, accreting sources whose isotropic X-ray luminosity exceeds the Eddington limit of a $\sim10\solarM$ \ac{BH}, that is $L_\mathrm{X}\gtrsim10^{39}$\,erg\,s$^{-1}$ \citep[see the reviews from][]{Kaaret2017,Fabrika2021,King2023_review,Pinto2023}. First detected by the \textit{Einstein} mission \citep{Fabbiano1989}, they are found outside the nuclear regions of their host galaxies, excluding the possibility of accreting supermassive \acp{BH}. % \acp{SMBH}. 

Initially interpreted as intermediate-mass \acp{BH} (M $\simeq10^2-10^5\solarM$) accreting at sub-Eddington rates \citep[see e.g.][]{ColbertMushotzky1999}, from the early 2000s, more and more pieces of evidence pointed towards an interpretation of \aclp{XRB} accreting at super-Eddington rates \citep{King2001,Poutanen2007,Zampieri2009}. Population studies found that \acp{ULX} are preferentially located in or near star-formation regions \citep{Colbert2004,Mineo2013} and therefore \acp{ULX} would represent the high-luminosity tail of the high-mass X-ray binary population \citep[see e.g.][]{Wiktorowicz2017}. Their energy spectra show significant differences with respect to standard sub-Eddington \aclp{XRB}, hinting at a peculiar, ultraluminous state \citep{Gladstone2009,Sutton2013}.

The discovery of pulsations in the X-ray flux of M82\,X-2 \citep{Bachetti2014} proved that at least a fraction of \acp{ULX} are powered by accreting \acp{NS}. In the last 10 years, a few more \acp{PULX} have been identified \citep{Fuerst2016,Israel2017a,Israel2017b,Carpano2018,Sathyaprakash2019,RodriguezCastillo2020} and more candidates are known \citep[see Table 2 in][]{King2023_review}, showing that \acp{NS} \citep[M $<2.9$ M$_\odot$,][]{KalogeraBaym1996} can emit at luminosities up to 1000 times the Eddington limit. At the moment, only 6 confirmed extragalactic, persistent \acp{PULX} are known out of $\sim2000$ \acp{ULX} \citep{Kovlakas2020,Walton2022catalogue,Tranin2024_cat}. However, the similarities between the energy spectra of many \acp{ULX} with those of \acp{PULX} indicate that \acp{PULX} must represent a consistent fraction of the \ac{ULX} population \citep{Walton2018_hardcomponent}. 
This is also suggested by timing analysis arguments: \ac{PULX} spin signals are characterized by intense spin up rates ($-10^{-10}\,\mathrm{s}\,\mathrm{s}^{-1}\lesssim\dot{P}\lesssim-10^{-11}\,\mathrm{s}\,\mathrm{s}^{-1}$) and typically low pulsed fractions (PF$\sim5\%-15\%$), so that accelerated search techniques \citep{Ransom2002} and $\sim10000$ photons \citep{RodriguezCastillo2020} are required to detect the spin periods. Considering only \ac{ULX} observations with at least 10000 detected photons, \acp{PULX} represent a remarkable $\sim25\%$ of the \ac{ULX} population. The real percentage, however, is still uncertain.

Apart from the detection of their spin signals, there are very few tentative indications on the nature of the compact objects in \acp{ULX}: \acp{PULX} show hard spectra, suggesting that other \acp{PULX} might be hidden among \acp{ULX} with emission peaking at energies $\gtrsim2$\,keV \citep{Pintore2017,Gurpide2021a_ULXsample}; \ac{BH}- and \ac{NS}-\acp{ULX} follow different paths in a hardness-intensity diagram \citep{Gurpide2021a}; the spectrum of a non-pulsating source (M51\,ULX-8) shows a \ac{CRSF}, suggesting the presence of a \ac{NS} as the accretor \citep{Brightman2018Nat_M51ULX8}; some \acp{PULX} display \acp{QPO} at the mHz range \citep{Imbrogno2024}{, raising the possibility that other \acp{ULX} showing \acp{QPO} might also harbour \acp{NS}}.  All these pieces of evidence, together with the fact that \acp{ULX} often show long-term variability that hinders the acquisition of high-quality spectra and/or multiple observations, make a strong case for the use of new tools to help find and identify new candidate \acp{PULX}. 

\ac{AI} is a scientific field comprising approaches that, given an input, can generate diverse types of outputs based on human-defined objectives \citep{ISOAI}. 
It comprehends several sub-fields, including \ac{ML}, that use data to address diverse problems (e.g. classification and clustering). \ac{ML} algorithms can be supervised or unsupervised. The former learns the relationships between inputs and outputs using labelled datasets, while the latter can discover patterns without any a priori labelling. This ability of unsupervised algorithms is crucial for scarcely labelled data, which is a common scenario in astrophysics \citep[see, e.g.][]{Slijepcevic2022, PinciroliVago2023, Sen2022}. Among unsupervised algorithms, clustering groups data with similar characteristics into subsets, or clusters \citep{Michaud1997, jain1988algorithms}, to gain insights into the population structure. Clustering is effective in different fields, including medicine \citep{Alashwal2019, Stoitsas2022}, art \citep{Deng2019}, and astrophysics \citep{PrezDaz2024}, with the advantage of working with high-dimensional data \citep{Assent2012}, which is less interpretable by humans \citep{arxivint}.

This research employs the clustering algorithm \acp{GMM} \citep[][]{Reynolds2009, PrezDaz2024} to identify a new set of \acp{ULX} that share similar properties with known \acp{PULX}, despite the absence of detected pulsations. These \acp{ULX}, therefore, constitute the ideal candidate \acp{PULX} for future observations and advanced pulsation search techniques. The methodology used in this paper is described in Sect. \ref{sec:Methods}, the results are presented in Sect. \ref{sec:Results} and discussed in Sect. \ref{sec:discussion}, while the most promising candidates are presented in Sect. \ref{sec:CandidatePULXs}. We give the conclusions and future implications of this work in Sect. \ref{sec:Conclusions}.

\section{Methods}
\label{sec:Methods}

This section introduces the dataset, explains the quantitative metrics to evaluate the clustering output, and presents the components of the clustering pipeline and the use of \acp{DT} for explainability.

\subsection{Dataset}
\label{sec:dataset}

We based our dataset on the \ac{ULX} catalogue of \citet[][W22 hereafter]{Walton2022catalogue}, which cross-correlates the catalogues of the currently operating main X-ray missions (the \chandra\ Source Catalogue, \citealp[CSC2,][]{Evans2020_chandracat}; the \swift\ X-ray Telescope Point Source Catalog, \citealp[2SXPS,][]{Evans2020_SwiftCat}; the \xmm\ Serendipitous Source Catalogue \citealp[4XMM-DR10,][]{Webb2020_XMMCat}) with the HyperLEDA galaxy archive \citep{Makarov2014_HyperLEDA}. The W22 catalogue counts 1843 \ac{ULX} candidates in 951 different host galaxies.

Of all the aforementioned X-ray missions, \xmm{} has the largest effective area (1500\,cm$^2$ at 2\,keV) and shorter time resolution \citep[$\sim$73\,ms for the EPIC-pn in imaging mode,][]{Struder2001_pn}. Moreover, the mission has been employed to carry out extensive observational campaigns of \acp{ULX} \citep[e.g.][]{Pinto2020_NGC1313X-1,Pinto2021,RodriguezCastillo2020,Fuerst2021_P13,Furst2023,Belfiore2024,Imbrogno2024}, delivering the highest-quality spectra and temporal series of \acp{ULX}. Since this work aims at using simultaneously the spectral and temporal properties of \acp{ULX}, we exclusively relied on \xmm{} data.

To account for the most recent observations of \acp{ULX}, we cross-matched the W22 catalogue with the 4XMM-DR13, the latest release at the time of writing. We removed all data of the \acp{ULX} M82 X-1 and X-2, since \xmm\ is not able to resolve them\footnote{M82 X-1 and X-2 are separated only by 5$^{\prime\prime}$, \citep[see e.g.][]{Brightman2020_M82X1X2}. \xmm\ \ac{PSF} sizes are 6.6$^{\prime\prime}$, 6.0$^{\prime\prime}$ and 4.5$^{\prime\prime}$ for EPIC-pn, MOS1 and MOS2, respectively (FWHM at 1.5\,keV, \url{https://xmm-tools.cosmos.esa.int/external/xmm_user_support/documentation/uhb/xmmpsf.html}).}. We added the data of two known \acp{ULX}, NGC 1313 X-2 and IC 342 X-1, not present in the W22 catalogue most likely because of their large offset from the centre of the host galaxies \citep[e.g. NGC 1313 X-2 is $\sim7^\prime$ distant from the host galaxy,][]{Sathyaprakash2022_NGC1313X2pulse}. 
The Galactic transient \ac{PULX} Swift J0243.6+6124 \citep{Kennea2017ATel_SwiftJ0243,WilsonHodge2018_SwiftJ0243} and those in the Magellanic Clouds \citep[RX J0209.6-7427 and SMC X-3][]{Vasilopoulos2020_RXJ0209,Weng2017_SMCX3,Towsend2017_SMCX3,Tsygankov2017_SMCX3} are also not included in W22 and are excluded from the present sample since they have not exhibited super-Eddington luminosities over long period of time (from months to years).

In the final dataset, each instance corresponds to an individual observation, identified by its source and observational identification numbers (\texttt{4XMMSRCID} and \texttt{OBS\_ID}). Consequently, a single source may have multiple observations. A summary of all the attributes (parameters) used in this work is given in Table \ref{tab:parameters}. The maximum observed flux (\fpeak) and luminosity (\lpeak) were updated, accounting for the 4XMM-DR13. With respect to W22, we added one further parameter, that is the hardness ratio defined as: 
\begin{equation}
\centering
\mathrm{HR_{Hard}}=\frac{H}{H+S}
\end{equation}
where $H$ and $S$ are the fluxes in the 2.0--12.0\,keV and 0.5--2.0\,keV energy bands, respectively. To check a posteriori where the majority of \acp{PULX} is and the properties of the sources belonging to the same cluster, we added three flags: \texttt{PULX} if the source is a known \ac{PULX}, \texttt{Pulse} if pulsations have been found in that observation, and \texttt{QPO} if \acp{QPO} have been reported in the literature for that observation. We do not flag the \ac{ULX} NGC 7793 ULX-4 as \texttt{PULX} due to the weakness of its pulsed signal \citep{Quintin2021}. Similarly, the \acp{ULX} M51 ULX-8 and NGC 5474 X-1, which exhibit (tentative or confirmed) \ac{CRSF} \citep{Brightman2018Nat_M51ULX8, Atapin2024_NGC5474X1_CRSF}, but no pulsations, are not flagged as \texttt{PULX}. It is worth emphasising that the great majority of observations with \acp{QPO}, 19 out of 22, are not associated with known \acp{PULX}.

In the end, our dataset is made up of  640 sources for a total of 1769 observations, with 95 observations (from 6 sources) flagged as \texttt{PULX}, 31 observations (from the same 6 sources) as \texttt{Pulse} \citep[][]{Fuerst2016,Israel2017a,Israel2017b,Carpano2018,RodriguezCastillo2020,Sathyaprakash2022_NGC1313X2pulse}, and 22 observations (from 8 sources) as \texttt{QPO} \citep[][Imbrogno et al., in prep.]{Atapin2019,Dewangan2005,Majumder2023,Pasham2015,Rao2010,Strohmayer2007,Strohmayer2009,Strohmayer2011,Agrawal2015,Heil2009,Imbrogno2024}. \\

\begin{table*}[!ht]
\caption{The parameters of the dataset used for the \ac{ML} algorithm.}
\label{tab:parameters}
\begin{tabular}{p{0.12\textwidth}p{0.8\textwidth}}
\toprule
Parameter    & Description \\ 
\midrule 
\texttt{Name} & Name of the host galaxy.\\
\texttt{Dist} & Distance of the host galaxy, in Mpc.\\
\texttt{4XMMName} & Name of the source as in the 4XMM catalogue, based on the source coordinates in the J2000 reference system.\\
\texttt{4XMMSRCID} &  4XMM source identification number, unique for each source.\\
\texttt{OBS\_ID} & 4XMM observation identification number.\\
\texttt{Fpeak4XMM} & Highest flux reached by the source in the broad energy band [8] 0.2-12 keV.\\
\texttt{Lpeak4XMM} & Highest luminosity reached by the source in the 0.2-12 keV band.\\
\texttt{EP\_n\_FLUX} & 4XMM fluxes in different energies bands. The parameter \texttt{n} denotes the following bands: [1] 0.2--0.5\,keV; [2] 0.5--1.0\,keV; [3] 1.0--2.0\,keV; [4] 2.0--4.5\,keV; [5] 4.5--12\,keV; [8] 0.2--12\,keV (broadband); [9] 0.5--4.5\,keV. \\
\texttt{VAR\_FLAG} & 4XMM flag on the source variability, set T(rue) only if the source was detected as variable in at least one exposure.\\
\texttt{HRHard} & Hardness ratio in the energies bands (2.0--12.0)\,keV/(0.5--12.0)\,keV\\
\texttt{PULX} & Label for known \acp{PULX}.\\
\texttt{Pulse}  & Label for observations where pulsations are detected. \\
\texttt{QPO} & Label for observations where \acp{QPO} are detected.\\
\bottomrule
\end{tabular}
\end{table*}

\subsection{Metrics}
\label{sec:metrics}

We first define two partitions that cover the dataset: $P$ and $U$. $P$ contains all the observations of sources flagged as \texttt{PULX}, while $U$ contains the remaining ones. We also define two clusters: $C_P$ and $C_U$. $C_P$ is the cluster that will contain most of the observations of known \acp{PULX} and the candidates found by the algorithm, while $C_U$ will gather the remaining observations. We introduce two quantitative metrics to evaluate the performance of our algorithm with respect to known \acp{PULX}: \ac{PR} and \ac{UR}. \ac{PR} is the ratio between the number of known observations of \acp{PULX} assigned to $C_P$ and the number of total observations of known \acp{PULX} ($|P|$):
\begin{equation}
    \ac{PR} = \frac{\max_j(|C_j \cap P|)}{|P|} = \frac{|C_P \cap P|}{|P|}
\end{equation}
where $C_j$ indicates a cluster $j$ and $|P|$ indicates the cardinality (i.e. the number of elements) of the set $P$. Note that, by definition, \ac{PR} is always $\geq$0.5; a high \ac{PR} is desired as it indicates that \acp{PULX} have similar characteristics. In this work, we impose a minimum \ac{PR} of 0.99.
\ac{UR} is the ratio between the number of observations of \acp{ULX} with unknown compact objects not contained in the cluster with the most \acp{PULX} ($C_P$) and the total number of \ac{ULX} observations with an unknown compact object type ($|U|$):

\begin{equation}
    \ac{UR} = \frac{|C_U \cap U|}{|U|}
\end{equation}
A high \ac{UR} value indicates that most observations of sources with uncertain compact object nature have different characteristics than confirmed \acp{PULX}. In summary, \ac{PR} and \ac{UR} enable a balance between maximising the identification of \acp{PULX} and avoiding the uninformative scenario of classifying everything as a \ac{PULX}.

In addition, we employ two widely used metrics to cross-check the results: the \ac{KS} p-value to quantify the difference in the distributions of all the attributes in the two clusters \citep{doi:10.1080/01621459.1951.10500769}, as implemented in \texttt{scipy} \citep{Virtanen2020}, and the \acl{SI} \citep[\acs{SI},][]{Rousseeuw1987}, to assess the ability of our algorithm to create well-separated clusters. Both metrics are computed after having applied \texttt{QuantileTransformer}\footnote{\url{https://scikit-learn.org/stable/modules/generated/sklearn.preprocessing.QuantileTransformer.html}} with a uniform output distribution to the input data.

For the \acs{SI} specifically, we consider each parameter separately. For each parameter, let $i$ and $j$ be the values of that parameter in two observations (or points). $C_A$ is the cluster (here, $C_U$ or $C_P$) that contains $i$ and $C_B$ is the other cluster.
The average dissimilarity $a(i)$ of $i$ to all the other points in $C_A$ is:
\begin{equation}
    a(i) = \frac{1}{|C_A|-1}\sum_{j \in C_A, i \neq j} d(i, j)
\end{equation}
where $d(i,j)$ is the distance between the points $i$ and $j$, defined as the absolute difference between $i$ and $j$.
Similarly, the average dissimilarity of a point $i \in C_A$ to the points of $C_B$ is computed as:
\begin{equation}
    b(i) = \frac{1}{|C_B|}\sum_{j \in C_B} d(i, j) .
\end{equation}
The quantity $s(i)$ represents the silhouette value of a point $i$ and is bounded between $-1$ and $1$:
\begin{equation}
    s(i) = \frac{b(i) - a(i)}{max\{a(i), b(i)\}} .
\end{equation}
The \acs{SI} is defined as the mean of all the $s(i)$ values. Higher values of \acs{SI} denote more compact and better-separated clusters.

Finally, the probability  $p$  of an observation to belong to the \ac{PULX} cluster is interpreted as the probability of that \ac{ULX} to be a candidate \ac{PULX}. The $z$-score \citep{zscoreref} is defined as $z = \Phi^{-1}(1-p)$ with:
\begin{equation}
    \Phi(x) = \frac{1+\erf\left(\frac{x}{\sqrt{2}}\right)}{2}
\end{equation}
where $\Phi(x)$ is the cumulative distribution function of the Gaussian distribution and erf is the error function.

\subsection{Pipeline}
\label{sec:pipeline}

\begin{figure}
    \centering
    \resizebox{\linewidth}{!}{%
\begin{tikzpicture}
    \draw[draw=black, ultra thick, solid] (-6.00,3.00) rectangle (-2.00,1.00);
    \draw[draw=black, ultra thick, solid] (0.00,3.00) rectangle (4.00,1.00);
    \draw[draw=black, ultra thick, solid] (7.00,2.00) rectangle (11.00,0.00);
    \draw[draw=black, ultra thick, solid] (10.00,-1.00) rectangle (14.00,-3.00);
    \draw[draw=black, ultra thick, solid] (7.00,-4.00) rectangle (11.00,-6.00);
    \draw[draw=black, ultra thick, solid] (4.00,-5.00) rectangle (0.00,-7.00);
    \draw[draw=black, ultra thick, solid] (-2.00,-5.00) rectangle (-6.00,-7.00);
    \draw[draw=black, ultra thick, loosely dashed] (6.00,3.00) rectangle (15.00,-7.00);
    \draw[draw=black, -latex, ultra thick, solid] (-2.00,2.00) -- (0.00,2.00);
    \draw[draw=black, -latex, ultra thick, solid] (4.00,2.00) -- (6.00,2.00);
    \draw[draw=black, -latex, ultra thick, solid] (0.00,-6.00) -- (-2.00,-6.00);
    \draw[draw=black, -latex, ultra thick, solid] (6.00,-6.00) -- (4.00,-6.00);
    \draw[draw=black, -latex, ultra thick, solid] (12.00,-5.00) -- (11.00,-5.00);
    \draw[draw=black, -latex, ultra thick, solid] (12.00,1.00) -- (12.00,-1.00);
    \draw[draw=black, ultra thick, solid] (11.00,1.00) -- (12.00,1.00);
    \draw[draw=black, ultra thick, solid] (12.00,-3.00) -- (12.00,-5.00);
    
    \node[black, anchor=center, text width=3.5cm, align=center, font=\Large] at (-4.00,2.00) {Input data};
    \node[black, anchor=center, text width=3.5cm, align=center, font=\Large] at (2.00,2.00) {Hyperparameters generation (step 1)};
    \node[black, anchor=center, text width=3.5cm, align=center, font=\Large] at (9.00,1.00) {Scaling and transformations (step 2a)};
    \node[black, anchor=center, text width=3.5cm, align=center, font=\Large] at (12.00,-2.00) {Clustering (step 2b)};
    \node[black, anchor=center, text width=3.5cm, align=center, font=\Large] at (9.00,-5.00) {UR and PR computation (step 2c)};
    \node[black, anchor=center, text width=3.5cm, align=center, font=\Large] at (2.00,-6.00) {Best clusters selection (step 3)};
    \node[black, anchor=center, text width=3.5cm, align=center, font=\Large] at (-4.00,-6.00) {Explanation with DT (step 4)};

    \node[black, anchor=north east, font=\large] at (14.9,2.9) {for each hyperparameters combination};
\end{tikzpicture}
}

    \caption{The pipeline for clustering the data and explaining the output, given a subset of attributes and a minimum \ac{PR}. 
    }
    \label{fig:pipeline}
\end{figure}

Our classifier comes with a number of hyperparameters (parameters specific to the
algorithm, rather than to the data) that need to be tuned. In order to do that, we follow a pipeline, outlined in Fig. \ref{fig:pipeline} and hereafter explained in its main steps.

\begin{enumerate}
    \item Generate hyperparameter combinations
    \item For each hyperparameter combination:
    \begin{enumerate}
        \item Apply data pre-processing (scaling and transformations);
        \item Fit \ac{GMM} on the \acp{ULX} observations with unknown nature;
        \item Select an optimal probability threshold on the output of \ac{GMM} on confirmed \ac{PULX} observations to maximize the \ac{UR} for the given \ac{PR} threshold of 99\% (cf. Sect. \ref{sec:metrics}); 
    \end{enumerate}
    \item Select the best hyperparameters (i.e. the ones maximising \ac{UR});
    \item Extract a \acs{DT} that explains the clustering output.
\end{enumerate}
The following subsections describe the relevant steps in more detail.

\subsubsection{Scaling and transformations}
\label{sec:preprocessing}

\begin{table}[]
\caption{The scaling,  transformation and \ac{GMM} hyperparameters and their values.}
\label{tab:hyperpre}
\begin{tabular}{@{}ll@{}}
\toprule
{Hyperparameter}    & {Possible values} \\ \midrule
Scaling & \begin{tabular}[c]{@{}l@{}}MinMaxScaler, StandardScaler,\\ QuantileTransformer, PowerTransformer\end{tabular} \\ 
\texttt{use\_PCA}                        & True, False              \\
\texttt{use\_log} & True, False \\
Cov. type & full, tied, diag, spherical \\
\bottomrule
\end{tabular}
\end{table}

First, data are transformed based on the first three hyperparameters summarised in Table \ref{tab:hyperpre}, in the following order:

\begin{enumerate}
    \item Logarithmic scale on the fluxes and \lpeak{} (if the hyperparameter \texttt{use\_log} is true)
    \item Scaling, with the algorithms implemented in \texttt{sklearn}\footnote{\url{https://scikit-learn.org/stable/modules/classes.html\#module-sklearn.preprocessing} } \citep{scikit-learn}.
    \item \ac{PCA} (if the hyperparameter \texttt{use\_PCA} is True), with 8 components, to create new sets of uncorrelated variables, as shown in \citet{KroneMartins2013}. 
\end{enumerate}

Scaling consists of adjusting the range of the attributes so that they fall within a specific interval or follow a distribution. In this work, we use min-max scaling\footnote{\url{https://scikit-learn.org/stable/modules/generated/sklearn.preprocessing.MinMaxScaler.html}}, standard scaling\footnote{\url{https://scikit-learn.org/stable/modules/generated/sklearn.preprocessing.StandardScaler.html}}, quantile transform\footnote{\url{https://scikit-learn.org/stable/modules/generated/sklearn.preprocessing.QuantileTransformer.html}} and power transform\footnote{\url{https://scikit-learn.org/stable/modules/generated/sklearn.preprocessing.PowerTransformer.html}}. Only one scaling method is used at a time. It is consistently applied across all attributes, ensuring they have similar ranges or distributions. Without scaling, some features might dominate others due to differences in scale. \ac{PCA} is a technique used for dimensionality reduction and feature extraction. It transforms the original set of attributes into a new set of uncorrelated attributes (the principal components) that can capture most of the variance in the original dataset.

\subsubsection{Clustering}
\label{sec:clustering}

We used \ac{GMM} on the scaled and transformed data to group observations with similar properties together, with the purpose of finding candidate \acp{PULX} similar to the known ones. \ac{GMM} is a probabilistic approach that can represent subpopulations with distinct characteristics inside a bigger population. 
\ac{GMM} is a soft clustering algorithm: it assigns to each data point the probability to belong to each cluster, in contrast to other algorithms which assign cluster labels \citep[e.g. k-means;][]{macqueen1967some}, and relies on a probabilistic framework \citep{reynolds2009gaussian}. 

\ac{GMM} outputs the weighted sum of $M$ Gaussian component densities \citep{Reynolds2009}:

\begin{equation}
    p(x|\lambda) = \sum_{c=1}^{M} w_c g(x | \mu_c, \Sigma_c)
\end{equation}

where $x \in \mathbb R^d$ is the input of the model, $d$ is the number of input parameters, $g(x | \mu_c, \Sigma_c)$ is a Gaussian (i.e. a component) with mean vector $\mu_c \in \mathbb R^d$ and covariance matrix $\Sigma_c \in \mathbb R^{d\times d}$, and $w_c$ is the weight associated with each component. $\lambda = \{w_c, \mu_c, \Sigma_c\}$, with $c \in {1, ..., M}$, is the collection of the \ac{GMM} weights, mean vectors, and covariance matrices. The weights $w_c$ can be interpreted as the proportion of the data associated with each component $c$.

First, we use the \texttt{BayesianGaussianMixture} implementation of \ac{GMM}\footnote{\url{https://scikit-learn.org/stable/modules/generated/sklearn.mixture.BayesianGaussianMixture.html}} with two components (one per cluster), where the covariance type (full, tied, diag, or spherical, see Table \ref{tab:hyperpre}) is the only variable hyperparameter and the priors for $\lambda$ are the default ones. \ac{GMM} is fitted only on the observations that are not associated with known \acp{PULX}. Then, \ac{GMM} outputs the probabilities of each observation (in the entire dataset) to belong to the two clusters. Since the sum of the two probabilities for each observation is 1, we use only the probability to belong to the \ac{PULX} cluster, hereafter. The probability threshold set at step 2(c) guarantees a minimum \ac{PR} and maximises \ac{UR}. Step 3 outputs the best clustering among all hyperparameter choices, that is, the one with the highest \ac{UR} among those selected at step 2.

\subsubsection{Explainability}
\label{sec:explainability}

Binary \acsp{DT} are non-parametric supervised learning algorithms used for classification and regression. They can model decisions by inferring rules from input features \citep{Kingsford2008}. Such rules correspond to axis-aligned cuts in an n-dimensional space that separate values below and above a specific threshold for each input feature. 
We use \acsp{DT} to approximate the best \ac{GMM} outputs. To this end, \ac{GMM} clusters are the target labels during the training of the \acsp{DT}. 

A \acs{DT} is read from top to bottom. Each condition leads to two branches (when the condition is met, on the left, and when it is not met, on the right), so the leaves (i.e. the nodes without children) are not associated with conditions. Each node contains five pieces of information:
\begin{itemize}
    \item A condition used to determine the child nodes at the subsequent level;
    \item The \texttt{gini} value, a decimal number between 0 and 1 that represents the impurity of a node, where low values indicate that all the samples belong to the same cluster (note that for two clusters the \texttt{gini} value can go from 0 to 0.5, with the latter corresponding to the maximum impurity level). 
    \item The \texttt{samples} value, which indicates the number of samples before applying the condition;
    \item \texttt{value}, that indicates the number of samples belonging to each of the clusters generated by \ac{GMM} (in the format $[$\ac{ULX}, \ac{PULX}$]$) before applying the condition (i.e. considering only the region bounded by the preceding conditions in the tree);
    \item The \texttt{class}, which indicates the predicted cluster to which most of the observations in that node belong.
\end{itemize}

\acsp{DT} explainability depends on their leaves depth (i.e. the number of steps from the top node, or root, of the tree), as shown in \citet{BLANCOJUSTICIA2020105532}. In this work, we use shallow trees (up to a depth of 4) as they describe the obtained clusters well, are easier to interpret, and provide a set of measurable parameters with a physical interpretation. Table \ref{tab:parameters} lists and describes the parameters considered in this work.

\section{Results}
\label{sec:Results}

\begin{table}[]
\centering
\caption{The maximum \ac{UR} achieved when considering or neglecting \fpeak{} or \lpeak{} and a minimum \ac{PR} of 0.99.}
\label{tab:results-ur-pr}
\begin{tabular}{@{}cccc@{}}
\toprule
\fpeak{} & \lpeak{}  & \ac{UR} \\ \midrule
 &  & 0.10 \\
 & \cmark & 0.11 \\
\cmark &  & 0.79 \\
\cmark & \cmark & 0.79 \\ \bottomrule
\end{tabular}
\end{table}

We run the algorithm for 4 different cases: with both \fpeak{} and \lpeak{}, with only \fpeak{}, with only \lpeak{}, and without \fpeak{} and \lpeak{}. Table \ref{tab:results-ur-pr} shows the best \ac{UR} for each case when \ac{PR} is at least 0.99. Keeping \fpeak{} leads to the best results, while \lpeak{} has a negligible effect, independent of the presence of \fpeak{}. In this case, the best scaler is MinMaxScaler. Avoiding the logarithmic transformation yields better results, omitting \ac{PCA} is preferable, and the best \ac{GMM} covariance type is "full".

\begin{figure*}[htbp]
    \centering
    \begin{subfigure}[b]{0.45\textwidth}
        \centering
        \includegraphics[width=\linewidth]{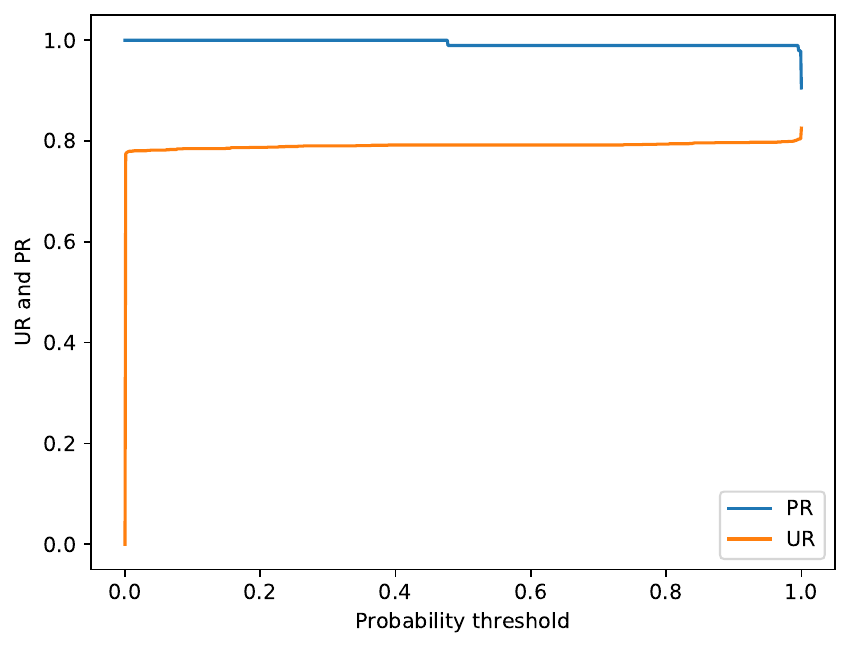}
        \label{fig:ur-pr-1-1}
    \end{subfigure}
    \hfill
    \begin{subfigure}[b]{0.45\textwidth}
        \centering
        \includegraphics[width=\linewidth]{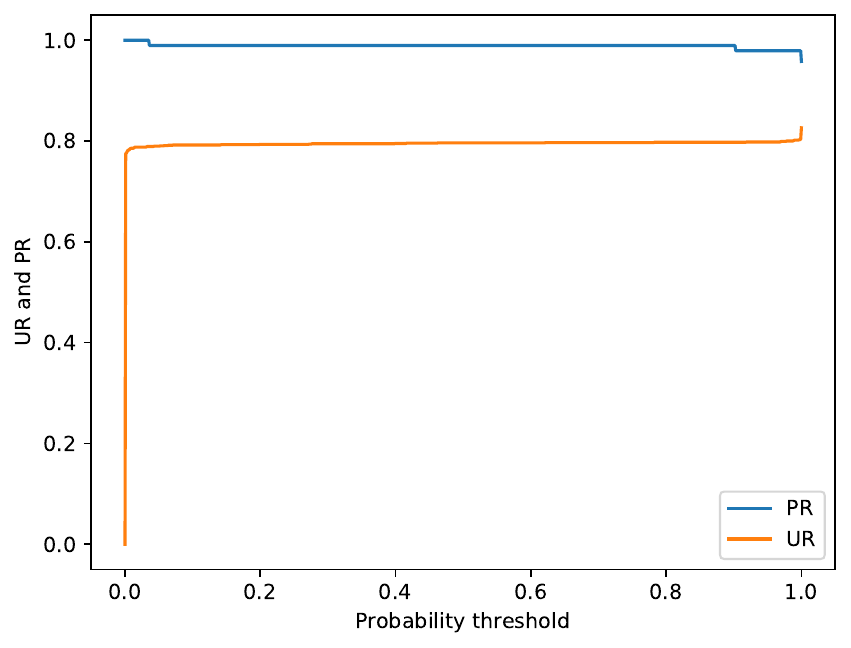}
        \label{fig:ur-pr-1-0}
    \end{subfigure}
    \caption{\ac{UR} and \ac{PR} as a function of the probability threshold when considering both \fpeak\ and \lpeak{} (left panel) or only \fpeak\ (right panel), that is, the two best cases reported in Table \ref{tab:results-ur-pr}.} 
    \label{fig:ur-pr}
\end{figure*}

Fig. \ref{fig:ur-pr} presents the \ac{UR}-\ac{PR} plots as a function of the probability threshold for the two best cases listed in Table \ref{tab:results-ur-pr}. Whenever we keep \fpeak{}, the variations in \ac{PR} are smaller as the probability threshold varies, and the variations in \ac{UR} are only observed when the probability threshold is 0 (i.e. all the observations are clustered together). The low variability and the high \ac{PR} and \ac{UR} values show that the pipeline can well separate the two clusters when keeping \fpeak{} and that the algorithm is robust to threshold variations. The curves are not monotonous because, by definition, \ac{PR} is computed on the cluster containing most of the known \acp{PULX}  and its value can never be lower than 0.5 by definition. When most of the known \acp{PULX} are below the probability threshold, the \acp{PULX} cluster ($C_P$) is the one with low probabilities (to be intended as probabilities of not observing a \ac{PULX}).

From now on, we focus on the cases with \fpeak{} and without \lpeak{} and a minimum \ac{PR} > 0.99. All the parameters resulted in \ac{KS} p-values $<10^{-5}$, while the \acs{SI} values are reported in Table \ref{tab:ks}. Bearing in mind that the \acs{SI} values between 0 and +1 denote that the clustering algorithm achieved a reliable assignment of the objects to their clusters, we found that both the \ac{KS} p-values and \acs{SI} values show that the data in the two clusters belong to significantly different distributions even when considering each variable separately. Moreover, the average distance between points in the same cluster is smaller than the distance between points in different clusters, even considering each variable separately.

\begin{table}[]
\centering
\caption{The results for the \acs{SI} for each variable in the case of \fpeak{} only. 
}
\label{tab:ks}
\begin{tabular}{@{}ccc@{}}
\toprule
 & \acl{SI} \\ 
 \midrule
Fpeak4XMM & 0.426 \\
EP\_1\_FLUX & 0.297 \\
EP\_2\_FLUX & 0.342 \\
EP\_3\_FLUX & 0.368 \\
EP\_4\_FLUX & 0.369 \\
EP\_5\_FLUX & 0.362 \\
EP\_8\_FLUX & 0.401 \\
EP\_9\_FLUX & 0.390 \\
VAR\_FLAG & 0.547 \\
HRHard & 0.039 \\
\bottomrule
\end{tabular}%
\end{table}

\begin{figure*}
    \centering
    \includegraphics[width=\textwidth]{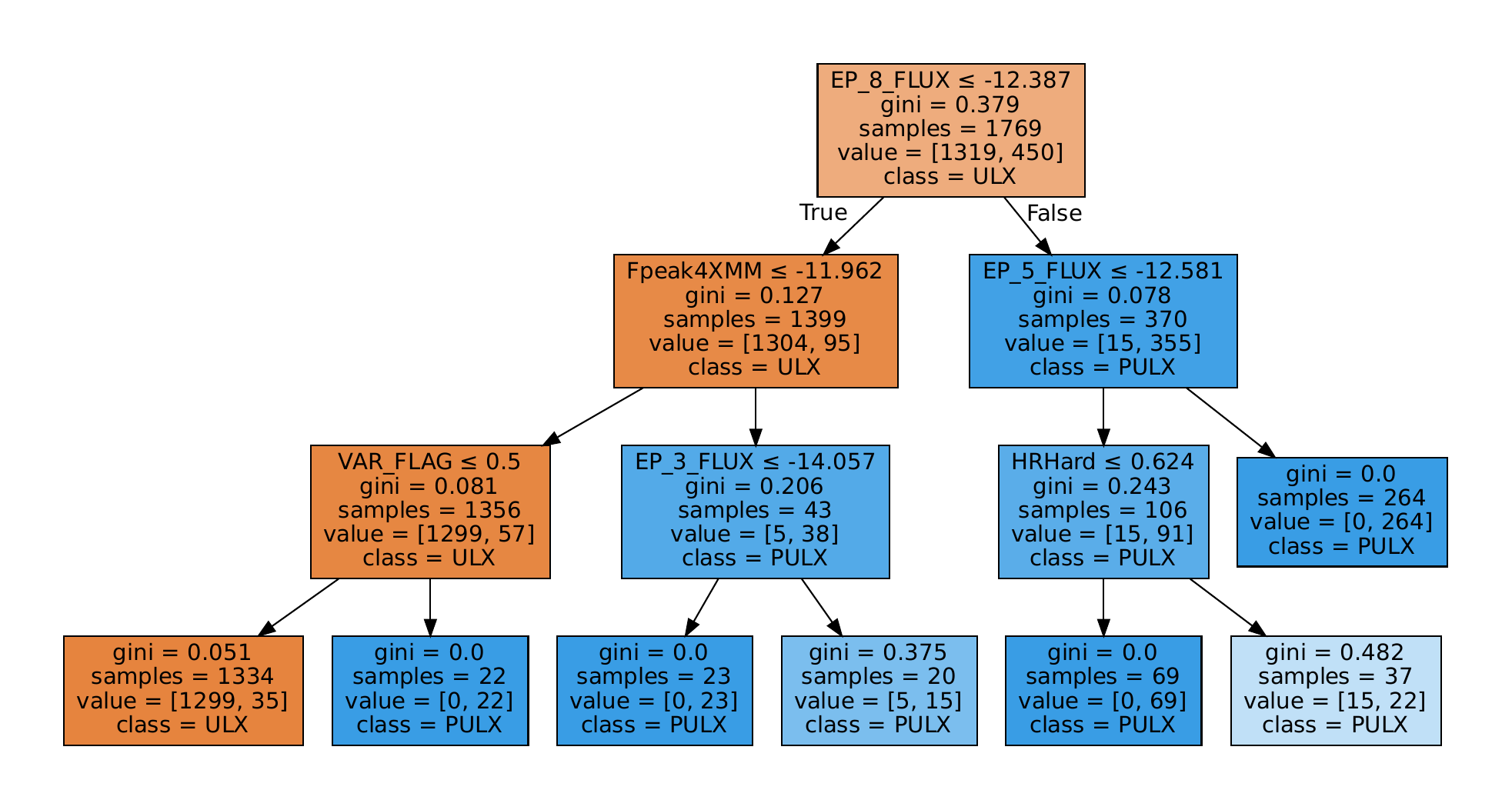}
    \caption{The \acs{DT} that describes the output of \ac{GMM} when keeping \fpeak{} and neglecting \lpeak, for \ac{PR} = 0.99. Orange nodes correspond to a majority of \acp{ULX}, while blue nodes correspond to a majority of \acp{PULX}.}
    \label{fig:dt}
\end{figure*}

Fig. \ref{fig:dt} shows the best \acs{DT}, which explains \ac{GMM} output for a minimum \ac{PR} of 0.99. The \acs{DT} uses six variables and shows that most of the new candidate \acp{PULX} are, in general, characterised by high fluxes (\logflux{} > -12.387 in the \EPVIIIFLUX{} energy band). Among them, 37 observations (22 sources) are classified as \acp{PULX} have a high broadband flux, a low flux in the \EPVFLUX{} energy band, and a high \texttt{HRHard}, but \texttt{gini} > 0 indicates that some observations not clustered with known \acp{PULX} also have the same characteristics. A smaller sample of candidate \acp{PULX} (43 observations of 10 sources) is also characterised by a high \fpeak{} (log(\fpeak{}) > -11.962). 
Finally, 22 observations (16 sources) with low fluxes, a low \fpeak{} and flagged as variable are also classified as \acp{PULX}.

\begin{figure*}[htbp]
    \centering
    \begin{subfigure}[b]{0.45\textwidth}
        \centering
        \includegraphics[width=\linewidth]{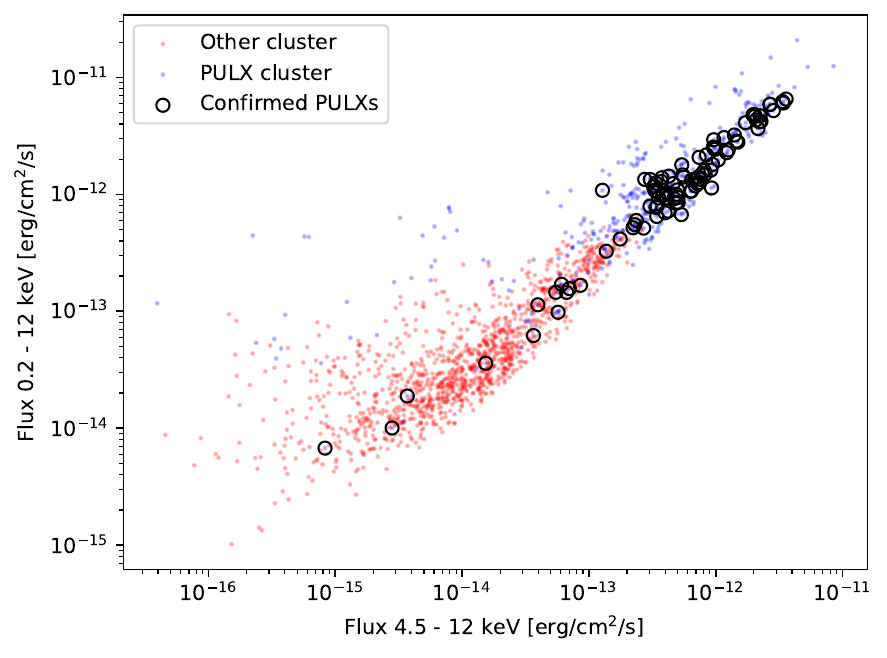}
        \label{fig:ep_5_ep_8}
    \end{subfigure}
    \hfill
    \begin{subfigure}[b]{0.45\textwidth}
        \centering
        \includegraphics[width=\linewidth]{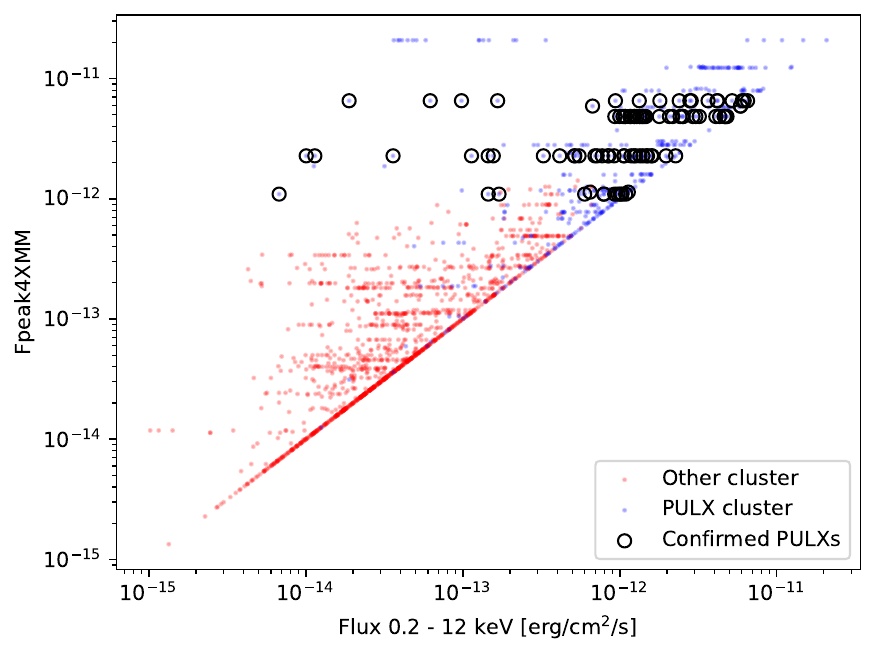}
        \label{fig:ep_8_fpeak}
    \end{subfigure}

    \caption{2D representations of the two clusters considering two pairs of parameters. The blue dots are the observations belonging to the $C_P$ cluster, the red dots are those belonging to the $C_U$ cluster, and the black empty circles flag the known \acp{PULX}. }
    \label{fig:scatter}
\end{figure*}

\begin{figure*}[htbp]
    \centering
    \begin{subfigure}[b]{0.45\textwidth}
        \centering
        \includegraphics[width=\linewidth]{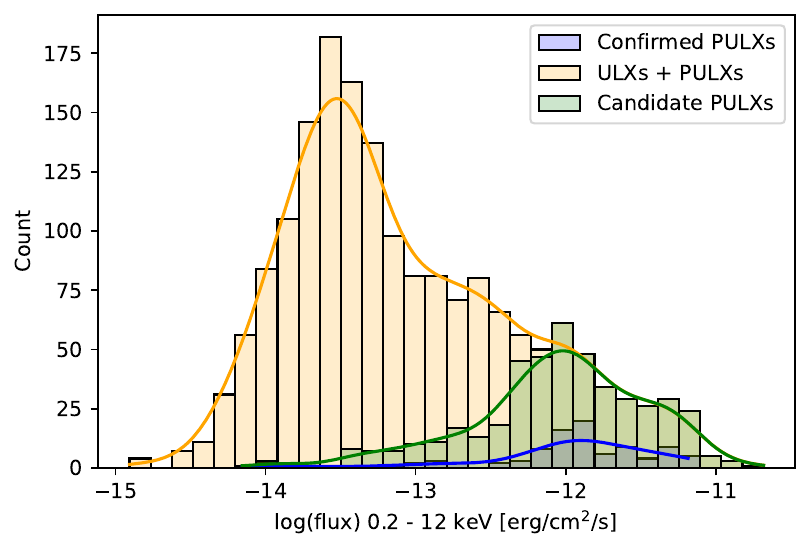}
        \label{fig:dist8}
    \end{subfigure}
    \hfill
    \begin{subfigure}[b]{0.45\textwidth}
        \centering
        \includegraphics[width=\linewidth]{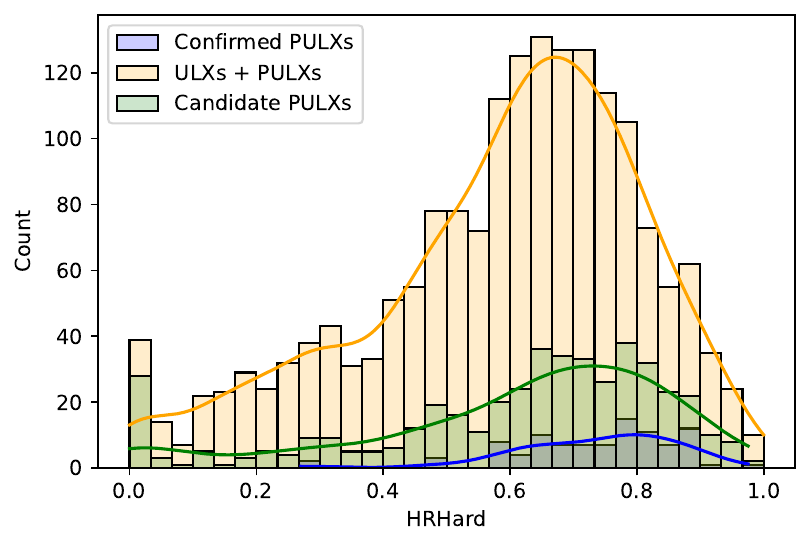}
        \label{fig:dist5}
    \end{subfigure}
    \caption{The distributions of two relevant parameters when neglecting only \lpeak{} for the whole sample (orange), the confirmed \acp{PULX} (blue) and the candidate \acp{PULX} (green).
    }
    \label{fig:distributions}
\end{figure*}

Fig. \ref{fig:scatter} represents the distribution of the two clusters, $C_P$ and $C_U$, using two examples of parameter combinations. The blue points indicate the observations belonging to $C_P$ (i.e. candidate and known PULXs), the red points those to $C_U$, and the black circles indicate the observations of known \acp{PULX} (including those where pulsations are not detected). In the case of \ac{PR} > 0.99, all the observations of known \acp{PULX} are retrieved. The plots also show that the two clusters are not well-separated in two dimensions and highlight the need to consider multiple parameters. Comparing the plots and the \acs{DT} shows that couples of parameters capture only a partial view of the \acs{DT}. 
Fig. \ref{fig:distributions} presents the distribution of two parameters, the broadband flux and \texttt{HRHard}. The cluster with candidate \acp{PULX} follows the same distribution as known \acp{PULX}, with a peak at high values of fluxes, while the \texttt{HRHard} distribution is similar for all three groups.

\begin{figure}
    \centering
    \includegraphics[width=0.95\linewidth]{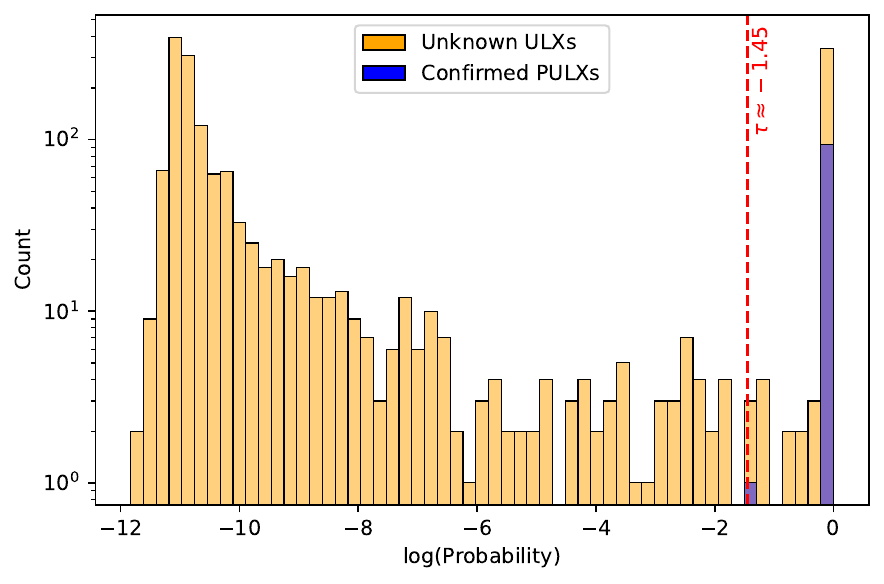}
    \caption{Distribution of \acp{PULX} with respect to the \ac{GMM} probability, in log-log scale.}
    \label{fig:distribution_probability}
\end{figure}

Fig. \ref{fig:distribution_probability} presents the distribution of the \acp{ULX} with respect to the probability determined by \ac{GMM}. The orange bars refer to \acp{ULX} with no detected pulsations, and the blue bars refer to \acp{PULX}. The red dashed line is the threshold $\tau$ (on the logarithm of the probability) corresponding to the confirmed \ac{PULX} with the lowest probability. The orange bars above the $\tau$ are 85 new candidate \acp{PULX} (the complete list is reported in Table \ref{tab:complete-table}). Neglecting the \ac{PULX} observation with the lowest probability, $\tau \approx -0.04$. As a result, 81 out of the 85 new candidate \acp{PULX} would still be identified as such. Considering all the 85 candidate \acp{PULX}, 52 sources have all the observations (276 in total) in the \ac{PULX} cluster, and 33 have observations (206 in total) in both clusters (see Table \ref{tab:distribution-candidates}). Moreover, 549 sources do not have observations (1192 in total) in the \ac{PULX} cluster. Overall, the candidate \acp{PULX} with all the observations in the \ac{PULX} cluster are $\approx 8.6$ times more than the confirmed \acp{PULX}, and only $\approx 5\%$ of the sources (corresponding to $\approx 12\%$ of the observations) are divided into multiple clusters.

\begin{figure}
    \centering
    \includegraphics[width=\linewidth]{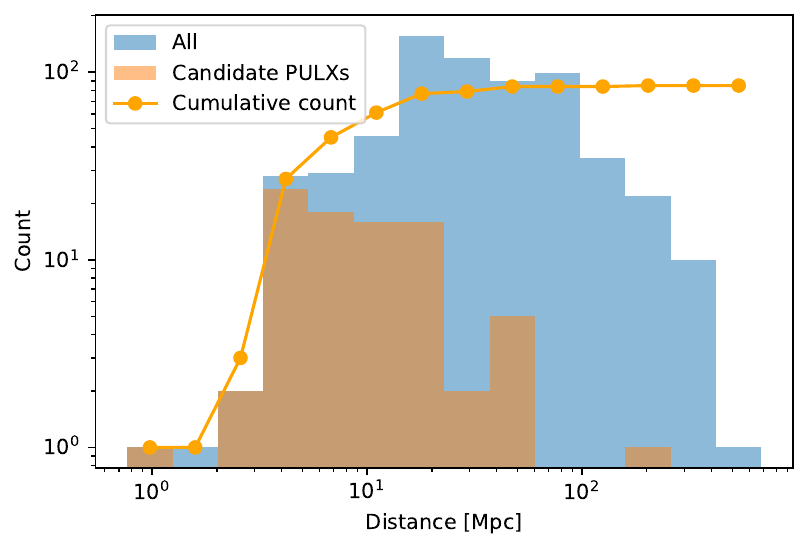}
    \caption{Distribution of the candidate \acp{PULX} (orange) and of the whole observations (cyan) as a function of distance and the cumulative count of candidate \acp{PULX}. Most of them have distances ranging between 2 and 20 Mpc. The candidate \acp{PULX} are obtained for a minimum \ac{PR} > 0.99.
}
    \label{fig:distances_candidates}
\end{figure}

Fig. \ref{fig:distance} presents the variation of the \ac{UR} as a function of the source distance, for \ac{PR} > 0.99, where for a given distance, all sources within that distance are considered. The \ac{UR} is high and the variations in \ac{UR} are small ($\approx$ 10\%) when keeping \fpeak{}. Finally, Fig. \ref{fig:distances_candidates} shows the distribution of the candidate \acp{PULX}' distances for \ac{PR} > 0.99. Most candidates are found between 2 and 40 Mpc. Moreover, the distribution suggests that closer sources are slightly more likely to be selected as candidate \acp{PULX}.

\section{Discussion}
\label{sec:discussion}

The goal of this work is to highlight archival \xmm{} observations of \acp{ULX} with similar characteristics to those of known \acp{PULX} in order to collect a sample of candidate \acp{PULX} based on the largest possible range of information for each of them. The novelty of this work is to use \ac{ML} methodologies, instead of more traditional techniques based on spectral and temporal analysis. In particular, we used \ac{GMM} to cluster all data in two groups: one that collects the majority of known \acp{PULX} ($C_P$) and one where the majority of \acp{ULX} have a compact object of unknown nature ($C_U$). We used two parameters for the maximum flux and luminosity reached by the source (parameters \fpeak\ and \lpeak, Table \ref{tab:results-ur-pr}). Only the former had an impact on our classifier. 

In particular, for a \ac{PR} > 0.99 (i.e. all known \acp{PULX} are correctly associated to the $C_P$ cluster) the best \acs{DT} (see Fig.  \ref{fig:dt}) can be reduced to just three main criteria to assign an observation to the \ac{PULX} class: a) a broadband flux $\gtrsim4\times10^{-13}$ \fluxunit{}, b) a maximum observed flux \fpeak{} $\gtrsim1\times10^{-12}$ \fluxunit{}, and c) the presence of intra-observational fractional variability (\texttt{VAR\_FLAG} $\geq$ 0.5). Based on this classification, the majority of observations in the $C_P$ cluster are those with the highest values of broadband fluxes (see also Fig.\ref{fig:distributions}, left panel) and \fpeak{}. Inside the $C_P$ cluster, together with the known \acp{PULX}, there are 355 additional observations for a total of 85 unique sources (Table \ref{tab:complete-table}) that constitute the candidate \acp{PULX} of this work.

Interestingly, parameters such as the hardness ratios or the fluxes in other energy bands do not seem to be decisive, as the \acs{DT} shows that most of the instances are classified as \acp{PULX} independently from them (Fig.\ref{fig:dt}): these parameters contribute to further refining the cluster membership without being discriminant. As a matter of fact, while the majority of instances assigned to the \ac{PULX} cluster are sources caught in a hard state, some of them have softer spectra and are still classified as candidate \acp{PULX}, including those of known \acp{PULX}. In particular, this is true for the only observation of M51 ULX-7 at the minimum of the super-orbital period, {classified as candidate \ac{PULX} \citep[ObsId.0830191401, with a flux of $2\times10^{-14}$\fluxunit, where no pulsations are detected due to poor statistics;][]{RodriguezCastillo2020}.} A similar result holds for the observation of NGC 5907 ULX-1 when the source was at $\sim$3$\times$10$^{-14}$\fluxunit\ and likely transitioning to a propeller regime \citep[ObsId. 0824320501;][]{Furst2023}. This is likely due to the parameter \fpeak{} imparting to the \ac{ML} algorithm that the source is related to another observation with a higher flux. 
To check this hypothesis, we kept the Gaussian distributions fitted on the whole sample of \acp{ULX} (without the known PULXs) using \ac{GMM} and the threshold determined on the original data. Then, we replaced the \fpeak{} values of the observations of known \acp{PULX} with the corresponding fluxes and computed the probabilities associated with the modified observations (this procedure does not modify the two initial Gaussians, but only affects where the points, corresponding to known PULXs, are positioned within the Gaussians). As a result, 13 observations of known \acp{PULX} ($\sim14$\%) are not correctly classified in the \ac{PULX} cluster, as their fluxes are particularly low. Of them, 12 observations (including ObsId.0830191401 of M51 ULX-7) do not show pulsations, and only one (ObsId.0804090401) does: this observation corresponds to the observation of NGC 5907 ULX-1 with the lowest flux, where the spin signal is seen \citep{Furst2023}. This finding suggests that the parameter \fpeak{} is important for a correct classification of low-flux observations, emphasising the need for multiple observations of \acp{ULX}, but it is not crucial, as most of the \ac{PULX} observations are retrieved even without using \fpeak{} at testing time. Moreover, there are candidate \acp{PULX} with a low \fpeak{} ($\approx 10^{-14}$ \fluxunit{}) and other \acp{ULX} with a high \fpeak{} ($\approx 10^{-12}$ \fluxunit{}).

\begin{figure}
    \centering
    \includegraphics[width=\linewidth]{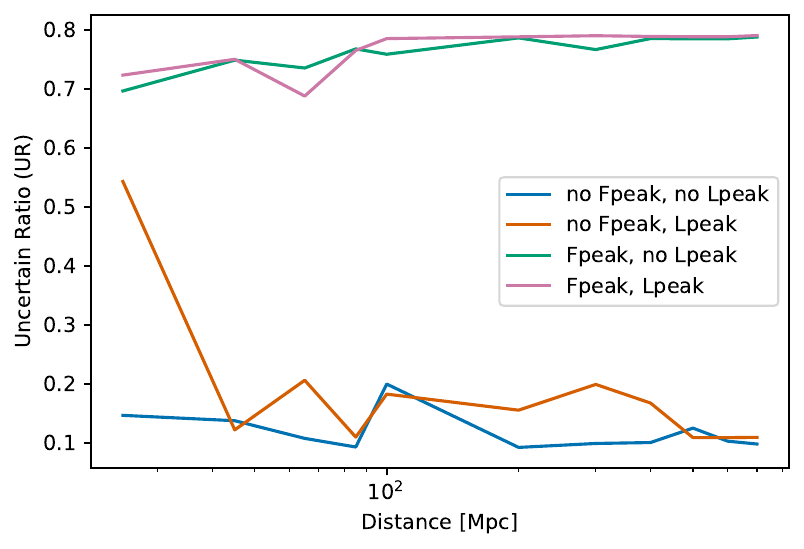}
    \caption{\ac{UR} as a function of the source distances, for different combinations of \fpeak{} and \lpeak{} and assuming \ac{PR} of 0.99. 
    }
\label{fig:distance}
\end{figure}

To further assess the robustness of our approach, we retain the fitted \ac{GMM} and analyse the impact of removing either a single known \ac{PULX} observation or all observations associated with a given \ac{PULX} on the \ac{PR}. Since \ac{GMM} is fitted only on the sources with unknown nature, removing \ac{PULX} observations affects only the threshold selection:
\begin{itemize}
    \item {Removing a single observation}: this affects the \ac{PR} only when the lowest-probability observation is neglected. The average \ac{PR} is computed by removing each observation (one at a time), recomputing the \ac{PR} for the entire dataset after each removal, and then averaging these values. The resulting average \ac{PR} is  $\approx 99\%$.
    \item {Removing a source}: we remove one source at a time, recompute the \ac{PR} for the entire dataset after each removal, and then compute the average across all individual observations. The resulting average \ac{PR} is $\approx 98\%$. Computing the average across sources, instead of observations, leads to a \ac{PR} of $\approx 97\%$.
\end{itemize}

This result shows that the pipeline would retrieve observations that were not initially included in the dataset. The effectiveness in retrieving \ac{PULX} observations, even when neglecting all the other observations of the same source, also shows that the pipeline does not rely only on source information such as \fpeak.

Concerning the distance of the sources (see Fig.\,\ref{fig:distance}), including sources within $\sim$30\,Mpc ($\sim$60\,Mpc when considering both \fpeak{} and \lpeak{}) marginally influences the performances of the algorithm, with values of \ac{UR} slightly smaller (keeping \ac{PR} > 0.99), suggesting that a reduced number of sources is not sufficient to achieve the best configuration. At the same time, the greater the distance, the higher the chance of contamination of misclassified sources. This seems to have little to no effect on the output, as \ac{UR} is approximately constant at distances $\gtrsim$100\,Mpc (for the best case of \ac{PR} > 0.99 and with only \fpeak{}). To see whether the plateauing of the \ac{UR} metrics is due to the presence of contaminants would require knowing their ratio over the total sample, and this is beyond the goal of this paper (see also below). Alternatively, the number of sources at distances $\gtrsim$100\,Mpc might be too small to influence the metrics significantly. Regardless, we emphasise that this does not modify the main findings of the work since the greatest part of the new \acp{PULX} is within 25\,Mpc.

Finally, a by-product of this work is the possibility of verifying the hypothesis that \acp{QPO} observed in \acp{ULX} can be used to mark the presence of an accreting \ac{NS} \citep[][Imbrogno et al., in prep.]{Imbrogno2024}. We found that 19 observations out of 22 where \acp{QPO} have been reported in the literature (8 sources in total) are included in the sample of the 85 new candidate \acp{PULX} listed in $C_P$ (the remaining three observations are of M51 ULX-7). Though this finding is far from being conclusive, it is suggestive that known \acp{PULX} and \acp{ULX} with unknown accretors displaying \acp{QPO} share some similarity in the multi-dimensional phase space studied by the \ac{ML} algorithm.

\subsection{Biases and caveats}
\label{subsec:caveats}

As for all \ac{ML} algorithms, the goodness of the results is limited by the number and the quality of available observational data. This naturally introduces biases. First and foremost, not all sources have been observed the same number of times. In general, \acp{ULX} hosted in the galaxies with astrophysical objects of great interest to the community, including \acp{PULX}, have a higher number of observations (e.g. NGC 1313 X-1 and X-2 have all been observed more than 30 times). On the contrary, faint \acp{ULX} typically have a smaller number of observations, with the risk of missing out on potential high-flux phases. In fact, all candidate \acp{PULX} have the highest \fpeak{} values in the dataset (note that \fpeak{} is equal to the flux of the observation if a source has been observed only once). This highlights the necessity to observe the sources multiple times in order to catch them in different luminosity states, to determine the true \fpeak{} and correctly classify them as candidate \acp{PULX} or not. In particular, considering that pulsations have been detected in 31 out of 95 observations of known \acp{PULX}, assuming the same distribution for both known and unknown \acp{PULX}, we can infer that a \ac{ULX} needs to be observed at least three times to have good chances ($\approx$70\%) to detect pulsations. 

As mentioned above, another issue is the presence of contaminants, mostly foreground and background sources (e.g. \acsp{AGN}), identified as point-like X-ray sources and associated with the wrong host galaxy. It is hard to obtain a precise estimate of the number of contaminants in the sample, but it has been noted that it grows as a function of distance, from a rate of $\lesssim$2\% for nearby galaxies up to 70\% for a pure \acs{HLX} sample \citep[see][and ref. therein]{Tranin2024_cat}. Considering that usually these distant contaminants have low fluxes, it can be reasonably excluded that they fall among the candidate \acp{PULX} found in this work, and hence, the sample of candidates can be considered reasonably pure.

Lastly, several caveats in this work need to be borne in mind. First, all instances used for the clustering are directly taken from either the 4XMM-DR13 or the W22 catalogues. In particular, the values of the fluxes in the different energy bands are taken from the 4XMM catalogue, where they are derived by fitting all the spectra with the same absorbed power-law model, with fixed equivalent hydrogen column and photon index values \citep[$n_\mathrm{H}=3\times10^{20}$~ atoms cm$^{-2}$, $\Gamma=1.7$;][]{Webb2020_XMMCat}. This is not the best model for \acp{ULX} spectra, which are better fitted with a double-component model \citep[such as multi-temperature discs or advection-dominated discs, see, e.g.][]{Gurpide2021a}, and in some cases with a high-energy cutoff \citep[e.g.][]{Walton2018_hardcomponent}. Hence, the fluxes used in this work are to be considered as an approximation of the actual source fluxes. Additionally, these fluxes are not corrected for interstellar or local absorptions, and hence, their temporal evolution does not accurately represent the intrinsic behaviour of the sources.
Moreover, due to the use of a power-law model, the fluxes in different energy bands are not exactly independent instances. Nonetheless, all the parameters have undergone a \ac{PCA} investigation (cf. Sect.\,\ref{sec:preprocessing}), where 8 components have been selected for the clustering algorithm. We stress that only some of the available information has been passed to the \ac{ML} algorithm. In particular, all the fields containing the statistical uncertainties of the fluxes are not taken into account, therefore considering each value as a "perfect" measurement. This assumption likely introduces some unknown level of bias that cannot be addressed by the used \ac{ML} algorithm. 

The main limitation of many clustering algorithms is that the number of clusters needs to be specified in advance \citep{Acquaviva2023}: in our approach, we considered only two clusters, representative of two classes of \ac{BH}- and \ac{NS}-\acp{ULX}. An alternative approach would be to consider the number of clusters $N_C$ as a free parameter, build a cluster scheme for different values of $N_C$, and choose the one providing the best result (in terms of cost function). Though powerful, this approach is more reliable with a larger number of observations than those currently available and would, in any case, still leave doubts regarding the physical interpretation of a greater number of clusters.

\subsection{Candidate \acp{PULX}}
\label{sec:CandidatePULXs}

While a detailed study of the candidate \acp{PULX} is beyond the aim of the present project and will be addressed in future works, we highlight the properties of some of them. In the \ac{PULX} cluster, there are 355 observations of sources not previously identified as \acp{PULX}, for a total of 85 candidates (Tables \ref{tab:complete-table} and \ref{tab:distribution-candidates}). All the candidates occupy the upper right corner of the plots of Fig.\,\ref{fig:scatter}. Hereafter, we briefly discuss the most promising ones. 

The observations with the highest flux belong to M31 ULX-1, in the Andromeda Galaxy. The source reached its peak flux in 2009 ($2\times10^{-11}$ \fluxunit{}, corresponding to $1.3\times10^{39}$ erg\,s$^{-1}$ at a bona fide distance of 766\,kpc), then slowly decayed and was last detected by \xmm{} in 2014. It has been interpreted as a stellar-mass \ac{BH} in a low-mass X-ray binary temporarily undergoing (nearly) super-Eddington accretion \citep{Middleton2012_M31ULX-1,Kaur2012}. 

A few other candidates have already been proposed as \ac{NS}-\acp{ULX}. In particular, \citet{Quintin2021} already detected a pulsation from NGC 7793 ULX-4 during its only outburst at super-Eddington luminosity. Because the detection of the spin signal was only marginally significant (3.4$\sigma$, cf. Sect.\,\ref{sec:dataset}), we excluded it from the sample of known \acp{PULX} used in this work. The clustering algorithm recovered it among the candidate \acp{PULX}, strengthening its nature as \ac{NS}-\ac{ULX}. Other good candidates are NGC 4559 X7, where a candidate pulsation with a significance of $3.5\sigma$ was recently found \citep{Pintore2025}; M81 X-6, suggested to host a weakly magnetised \ac{NS} based on its spectral evolution \citep{Amato2023}; IC 342 X-1 and NGC 925 ULX-1, previously identified as candidate \acp{PULX} because of their hardness ratios similar to that of known \acp{PULX} \citep{Pintore2017, Pintore2018_NGC925}. 

The \ac{ML} algorithm classifies as candidate \acp{PULX} also a few sources interpreted in the literature as \ac{BH}-\acp{ULX}, such as Holmberg II X-1 \citep{Barra2024_HoIIX-1}, NGC 5240 X-1, NGC 5408 X-1 \citep[e.g.][]{Gurpide2021a}, and NGC 7793 P9 \citep{Hu2018}. Other sources that do not offer sufficient spectral or temporal indications on the nature of their compact objects are also classified as candidate \acp{PULX}, as NGC 1313 X-1 \citep[e.g.][]{Walton2020}, IC 342 X-2, and NGC 5240 X-1. This strengthens the idea that traditional spectral or temporal analyses alone might not be able to uncover all the aspects of the \ac{ULX} phenomenology.

We further note that, for some sources, not all observations are assigned to the same cluster (see Table \ref{tab:distribution-candidates}). This is, for example, the case of M51 ULX-8, which has only one \xmm{} observation (ObsID.0677980801), out of 11, associated with the \ac{PULX} cluster, likely due to the high flux and, correspondingly, high statistics of the given datasets. All the other \xmm{} observations of this source resulted in low $z$-scores, preventing them from being associated with the same \ac{PULX} cluster.

Finally, we carried out a preliminary timing analysis on the candidates by using the \texttt{HENaccelsearch} within the \texttt{HENDRICS}/\texttt{Stingray} software packages \citep{Bachetti2018,Huppenkothen2019_apj,Bachetti2024_stingray2}, which accounts only for a first period derivative.
We searched for pulsations from 0.1\,Hz to the Nyquist frequency. The search gave negative results. More detailed timing analyses with both the traditional and \ac{AI} evolutionary algorithms are needed in order to exhaustively study the new candidates \citep[as done in][]{Sacchi2024}.

\section{Conclusions and future perspectives}
\label{sec:Conclusions}

This work presents the results of the application of an unsupervised clustering pipeline to a database of \ac{ULX} observations to identify new candidate \acp{PULX} relying upon all the available spectral/timing properties often neglected in standard analysis. We show that \ac{GMM} determines for each observation the probability of belonging to the cluster containing the confirmed \acp{PULX}. Our pipeline assigns high probabilities to the observations of confirmed \acp{PULX} when the parameter \fpeak{} is taken into account. For this reason, the clustering effectively isolates candidate \acp{PULX} for a wide range of probability thresholds. In this case, retrieving at least 99\% of the confirmed \acp{PULX} leads to 85 new candidate \acp{PULX} from 355 observations.

Future improvements of this work will focus on performing systematic fits of all \ac{ULX} spectra with more appropriate models (cf. Sect. \ref{subsec:caveats}), allowing for better flux estimates in the different energy bands. The algorithm will be rerun on the updated database to obtain more robust results. Moreover, more reliable fluxes will lead to more precise luminosity estimates, so that the luminosities can also be employed as cluster parameters to better investigate the dependency of the results on the distance. The uncertainties on the best-fit parameters will also be taken into account (e.g. with bootstrapping), to increase the reliability of the results.

Another issue concerns the choice of the luminosity threshold of 10$^{39}$ \lumunit\ to select potential \acp{PULX}, as the mere existence of X-ray pulsars with luminosities in the $\approx$10$^{34}-$ few 10$^{38}$ \lumunit\ range clearly casts doubts on its validity. 
Hence, an improvement on the present work will include observations of X-ray sources in the 2$\times$10$^{38}-$10$^{39}$ \lumunit\ range, which are potential X-ray pulsars in the super-Eddington regime. The inclusion of less luminous X-ray binaries can also contribute to the discussion of \acp{ULX} being an extension of the high-mass X-ray binary and \acp{PULX} of the X-ray pulsar populations.

The proposed pipeline is part of a ``living'' project and will be applied to new \ac{ULX} observations when they are performed. For this purpose, we envision the adoption of continual learning principles \citep{10444954}.

\begin{acknowledgements}
The database used in this research has been derived from the ULX catalogue of \citet{Walton2022catalogue} and the 4XMM-DR13 \citet{Webb2020_XMMCat}, and it is available upon request. This study also uses observations obtained with \xmm, a European Space Agency (ESA) science mission with instruments and contributions directly funded by ESA Member States and National Aeronautics and Space Administration (NASA).
NOPV acknowledges support from INAF and CINECA for granting 125,000 core hours on the Leonardo supercomputer to carry out the project "Finding Pulsations with Evolutionary Algorithms". MI is supported by the AASS Ph.D. joint research programme between the University of Rome "Sapienza" and the University of Rome "Tor Vergata", with the collaboration of the National Institute of Astrophysics (INAF). RA and GLI acknowledge financial support from INAF through the grant ``INAF-Astronomy Fellowships in Italy 2022 - (GOG)''. GLI also acknowledges support from PRIN MUR SEAWIND (2022Y2T94C), which was funded by NextGenerationEU and INAF Grant BLOSSOM. KK is supported by a fellowship program at the Institute of Space Sciences (ICE-CSIC) funded by the program Unidad de Excelencia Mar\'ia de Maeztu CEX2020-001058-M.

\end{acknowledgements}

\begin{acronym}
\acro{AGN}{Active Galactic Nucleus}
\acroplural{AGN}[AGNs]{Active Galactic Nuclei}
\acro{AI}{Artificial Intelligence}
\acro{BH}{black hole}
\acro{CRSF}{cyclotron resonant scattering feature} 
\acro{DT}{Decision Tree}
\acro{GMM}{Gaussian Mixture Model}
\acro{HLX}{hyperluminous X-ray source}
\acro{IMBH}{intermediate-mass black hole}
\acro{KS}{Kolmogorov-Smirnov}
\acro{ML}{Machine Learning}
\acro{NS}{neutron star}
\acro{PCA}{Principal Component Analysis}
\acro{PR}{PULXs Ratio}
\acro{PSF}{point spread function}
\acro{PULX}{Pulsating ULX}
\acro{QPO}{quasi-periodic oscillation}
\acro{SI}{Silhouette Index}
\acro{SMBH}{supermassive black hole}
\acro{ULX}{Ultraluminous X-ray source}
\acro{UR}{Uncertain Ratio}
\acro{XRB}{X-ray binary}
\acroplural{XRB}[XRBs]{X-ray binaries}
\end{acronym}

\bibliographystyle{aa} 
\bibliography{biblio} 
\onecolumn

\begin{appendix}

\section{Observations of known \acp{PULX}}

\begin{longtable}[c]{@{}cccccccc@{}}
\caption{The predicted $z$-scores for all the known \acp{PULX} in the dataset, grouped by name.}
\label{tab:confirmed-table}\\
\toprule
\multirow{2}{*}{{OBS\_ID}} & \multirow{2}{*}{{Name}} & \multirow{2}{*}{{4XMMName}} & \multirow{2}{*}{{Pulsations}} & \multicolumn{4}{c}{{z-score}} \\* \cmidrule(l){5-8} 
 &  &  &  & None & Only \lpeak & Only \fpeak & Both \\* \midrule
\endfirsthead
\multicolumn{8}{c}%
{{\bfseries Table \thetable\ continued from previous page}} \\
\toprule
\multirow{2}{*}{{OBS\_ID}} & \multirow{2}{*}{{Name}} & \multirow{2}{*}{{4XMMName}} & \multirow{2}{*}{{Pulsations}} & \multicolumn{4}{c}{{z-score}} \\* \cmidrule(l){5-8} 
 &  &  &  & None & Only \lpeak & Only \fpeak & Both \\* \midrule
\endhead
\bottomrule
\endfoot
\endlastfoot
656780401 & NGC0300 ULX1 & 4XMMJ005504.8-374143 &  & 4.70 & 0.51 & \textgreater 8.00 & \textgreater 8.00 \\
791010101 & NGC0300 ULX1 & 4XMMJ005504.8-374143 & \checkmark & \textgreater 8.00 & \textgreater 8.00 & \textgreater 8.00 & \textgreater 8.00 \\
791010301 & NGC0300 ULX1 & 4XMMJ005504.8-374143 & \checkmark & \textgreater 8.00 & \textgreater 8.00 & \textgreater 8.00 & \textgreater 8.00 \\* \midrule
150280701 & NGC1313 PULX & 4XMMJ031747.4-663009 &  & 4.10 & -5.93 & \textgreater 8.00 & \textgreater 8.00 \\
205230501 & NGC1313 PULX & 4XMMJ031747.4-663009 & \checkmark & 5.37 & 1.62 & 5.22 & 4.90 \\* \midrule
106860101 & NGC1313 X-2 & 4XMMJ031822.1-663603 &  & 7.72 & 2.72 & \textgreater 8.00 & \textgreater 8.00 \\
150280101 & NGC1313  X-2 & 4XMMJ031822.1-663603 &  & \textgreater 8.00 & 2.70 & \textgreater 8.00 & \textgreater 8.00 \\
150280201 & NGC1313  X-2 & 4XMMJ031822.1-663603 &  & \textgreater 8.00 & 1.51 & \textgreater 8.00 & \textgreater 8.00 \\
150280301 & NGC1313  X-2 & 4XMMJ031822.1-663603 &  & \textgreater 8.00 & 2.07 & \textgreater 8.00 & \textgreater 8.00 \\
150280401 & NGC1313  X-2 & 4XMMJ031822.1-663603 &  & \textgreater 8.00 & 1.83 & \textgreater 8.00 & \textgreater 8.00 \\
150280501 & NGC1313  X-2 & 4XMMJ031822.1-663603 &  & \textgreater 8.00 & 2.92 & \textgreater 8.00 & \textgreater 8.00 \\
150280601 & NGC1313  X-2 & 4XMMJ031822.1-663603 &  & \textgreater 8.00 & 2.84 & \textgreater 8.00 & \textgreater 8.00 \\
150280701 & NGC1313  X-2 & 4XMMJ031822.1-663603 &  & 7.77 & 2.72 & \textgreater 8.00 & \textgreater 8.00 \\
150281101 & NGC1313  X-2 & 4XMMJ031822.1-663603 &  & 7.93 & 2.80 & \textgreater 8.00 & \textgreater 8.00 \\
205230201 & NGC1313  X-2 & 4XMMJ031822.1-663603 &  & 7.81 & 2.79 & \textgreater 8.00 & \textgreater 8.00 \\
205230301 & NGC1313  X-2 & 4XMMJ031822.1-663603 &  & \textgreater 8.00 & 1.97 & \textgreater 8.00 & \textgreater 8.00 \\
205230401 & NGC1313  X-2 & 4XMMJ031822.1-663603 &  & 7.45 & 2.61 & \textgreater 8.00 & \textgreater 8.00 \\
205230501 & NGC1313  X-2 & 4XMMJ031822.1-663603 &  & 7.53 & 2.65 & \textgreater 8.00 & \textgreater 8.00 \\
205230601 & NGC1313  X-2 & 4XMMJ031822.1-663603 &  & \textgreater 8.00 & 1.62 & \textgreater 8.00 & \textgreater 8.00 \\
301860101 & NGC1313  X-2 & 4XMMJ031822.1-663603 &  & \textgreater 8.00 & 1.58 & \textgreater 8.00 & \textgreater 8.00 \\
405090101 & NGC1313  X-2 & 4XMMJ031822.1-663603 &  & \textgreater 8.00 & \textgreater 8.00 & \textgreater 8.00 & \textgreater 8.00 \\
693850501 & NGC1313  X-2 & 4XMMJ031822.1-663603 &  & \textgreater 8.00 & \textgreater 8.00 & \textgreater 8.00 & \textgreater 8.00 \\
693851201 & NGC1313  X-2 & 4XMMJ031822.1-663603 &  & \textgreater 8.00 & 2.80 & \textgreater 8.00 & \textgreater 8.00 \\
722650101 & NGC1313  X-2 & 4XMMJ031822.1-663603 &  & 6.88 & 2.40 & \textgreater 8.00 & \textgreater 8.00 \\
742490101 & NGC1313  X-2 & 4XMMJ031822.1-663603 &  & 7.97 & 2.87 & \textgreater 8.00 & \textgreater 8.00 \\
742590301 & NGC1313  X-2 & 4XMMJ031822.1-663603 &  & \textgreater 8.00 & \textgreater 8.00 & \textgreater 8.00 & \textgreater 8.00 \\
764770101 & NGC1313  X-2 & 4XMMJ031822.1-663603 &  & 7.06 & 2.65 & \textgreater 8.00 & \textgreater 8.00 \\
764770401 & NGC1313  X-2 & 4XMMJ031822.1-663603 &  & \textgreater 8.00 & 2.88 & \textgreater 8.00 & \textgreater 8.00 \\
782310101 & NGC1313  X-2 & 4XMMJ031822.1-663603 &  & \textgreater 8.00 & \textgreater 8.00 & \textgreater 8.00 & \textgreater 8.00 \\
794580601 & NGC1313  X-2 & 4XMMJ031822.1-663603 &  & \textgreater 8.00 & 2.81 & \textgreater 8.00 & \textgreater 8.00 \\
803990101 & NGC1313  X-2 & 4XMMJ031822.1-663603 &  & \textgreater 8.00 & \textgreater 8.00 & \textgreater 8.00 & \textgreater 8.00 \\
803990201 & NGC1313  X-2 & 4XMMJ031822.1-663603 &  & \textgreater 8.00 & \textgreater 8.00 & \textgreater 8.00 & \textgreater 8.00 \\
803990301 & NGC1313  X-2 & 4XMMJ031822.1-663603 &  & \textgreater 8.00 & 2.77 & \textgreater 8.00 & \textgreater 8.00 \\
803990401 & NGC1313  X-2 & 4XMMJ031822.1-663603 & \checkmark & \textgreater 8.00 & \textgreater 8.00 & \textgreater 8.00 & \textgreater 8.00 \\
803990501 & NGC1313  X-2 & 4XMMJ031822.1-663603 &  & \textgreater 8.00 & \textgreater 8.00 & \textgreater 8.00 & \textgreater 8.00 \\
803990601 & NGC1313  X-2 & 4XMMJ031822.1-663603 & \checkmark & 7.79 & 2.78 & \textgreater 8.00 & \textgreater 8.00 \\
803990701 & NGC1313  X-2 & 4XMMJ031822.1-663603 &  & 7.92 & 2.77 & \textgreater 8.00 & \textgreater 8.00 \\* \midrule
112840201 & M51 ULX-7 & 4XMMJ133000.9+471343 &  & 7.28 & \textgreater 8.00 & \textgreater 8.00 & \textgreater 8.00 \\
212480801 & M51 ULX-7 & 4XMMJ133000.9+471343 & \checkmark & \textgreater 8.00 & \textgreater 8.00 & \textgreater 8.00 & \textgreater 8.00 \\
303420101 & M51 ULX-7 & 4XMMJ133000.9+471343 &  & \textgreater 8.00 & \textgreater 8.00 & \textgreater 8.00 & \textgreater 8.00 \\
303420201 & M51 ULX-7 & 4XMMJ133000.9+471343 &  & \textgreater 8.00 & \textgreater 8.00 & \textgreater 8.00 & \textgreater 8.00 \\
677980701 & M51 ULX-7 & 4XMMJ133000.9+471343 &  & \textgreater 8.00 & \textgreater 8.00 & \textgreater 8.00 & \textgreater 8.00 \\
677980801 & M51 ULX-7 & 4XMMJ133000.9+471343 &  & 2.76 & -1.05 & -1.81 & -0.06 \\
824450901 & M51 ULX-7 & 4XMMJ133000.9+471343 & \checkmark & 6.90 & 2.58 & 5.96 & 6.49 \\
830191401 & M51 ULX-7 & 4XMMJ133000.9+471343 &  & -3.24 & -5.06 & 1.29 & 2.64 \\
830191501 & M51 ULX-7 & 4XMMJ133000.9+471343 & \checkmark & 7.45 & 2.64 & \textgreater 8.00 & \textgreater 8.00 \\
830191601 & M51 ULX-7 & 4XMMJ133000.9+471343 &  & 7.42 & 2.64 & \textgreater 8.00 & \textgreater 8.00 \\
852030101 & M51 ULX-7 & 4XMMJ133000.9+471343 & \checkmark & 7.34 & 2.63 & \textgreater 8.00 & \textgreater 8.00 \\* \midrule
145190101 & NGC5907 ULX-1 & 4XMMJ151558.6+561810 &  & 6.94 & 0.65 & \textgreater 8.00 & \textgreater 8.00 \\
145190201 & NGC5907 ULX-1 & 4XMMJ151558.6+561810 & \checkmark & 7.27 & 0.41 & \textgreater 8.00 & \textgreater 8.00 \\
673920201 & NGC5907 ULX-1 & 4XMMJ151558.6+561810 &  & 5.18 & -3.89 & \textgreater 8.00 & \textgreater 8.00 \\
673920301 & NGC5907 ULX-1 & 4XMMJ151558.6+561810 &  & 5.32 & -0.61 & \textgreater 8.00 & \textgreater 8.00 \\
724810401 & NGC5907 ULX-1 & 4XMMJ151558.6+561810 &  & 5.64 & 0.89 & \textgreater 8.00 & \textgreater 8.00 \\
729561301 & NGC5907 ULX-1 & 4XMMJ151558.6+561810 & \checkmark & 7.62 & 0.69 & \textgreater 8.00 & \textgreater 8.00 \\
795712601 & NGC5907 ULX-1 & 4XMMJ151558.6+561810 &  & -4.06 & -6.22 & \textgreater 8.00 & \textgreater 8.00 \\
804090301 & NGC5907 ULX-1 & 4XMMJ151558.6+561810 & \checkmark & 4.94 & 0.94 & \textgreater 8.00 & 6.84 \\
804090401 & NGC5907 ULX-1 & 4XMMJ151558.6+561810 & \checkmark & 3.68 & 0.47 & \textgreater 8.00 & 7.56 \\
804090501 & NGC5907 ULX-1 & 4XMMJ151558.6+561810 &  & 0.64 & -3.26 & \textgreater 8.00 & \textgreater 8.00 \\
804090601 & NGC5907 ULX-1 & 4XMMJ151558.6+561810 &  & -0.07 & -0.47 & \textgreater 8.00 & \textgreater 8.00 \\
804090701 & NGC5907 ULX-1 & 4XMMJ151558.6+561810 &  & -2.68 & -5.23 & \textgreater 8.00 & \textgreater 8.00 \\
804090801 & NGC5907 ULX-1 & 4XMMJ151558.6+561810 &  & 6.45 & 0.32 & \textgreater 8.00 & \textgreater 8.00 \\
804090901 & NGC5907 ULX-1 & 4XMMJ151558.6+561810 &  & 6.18 & -0.08 & \textgreater 8.00 & \textgreater 8.00 \\
804091001 & NGC5907 ULX-1 & 4XMMJ151558.6+561810 &  & 6.22 & 0.04 & \textgreater 8.00 & \textgreater 8.00 \\
804091101 & NGC5907 ULX-1 & 4XMMJ151558.6+561810 &  & 5.90 & 0.11 & \textgreater 8.00 & \textgreater 8.00 \\
804091201 & NGC5907 ULX-1 & 4XMMJ151558.6+561810 &  & 5.81 & -0.35 & \textgreater 8.00 & \textgreater 8.00 \\
824320201 & NGC5907 ULX-1 & 4XMMJ151558.6+561810 & \checkmark & 6.79 & 0.65 & \textgreater 8.00 & \textgreater 8.00 \\
824320301 & NGC5907 ULX-1 & 4XMMJ151558.6+561810 & \checkmark & 6.67 & 0.37 & \textgreater 8.00 & \textgreater 8.00 \\
824320401 & NGC5907 ULX-1 & 4XMMJ151558.6+561810 &  & 6.14 & 0.17 & \textgreater 8.00 & \textgreater 8.00 \\
824320501 & NGC5907 ULX-1 & 4XMMJ151558.6+561810 &  & -3.43 & -3.41 & \textgreater 8.00 & \textgreater 8.00 \\
824320601 & NGC5907 ULX-1 & 4XMMJ151558.6+561810 & \checkmark & 6.82 & 0.47 & \textgreater 8.00 & \textgreater 8.00 \\
824320701 & NGC5907 ULX-1 & 4XMMJ151558.6+561810 &  & 0.32 & -1.35 & \textgreater 8.00 & \textgreater 8.00 \\
851180701 & NGC5907 ULX-1 & 4XMMJ151558.6+561810 &  & 5.00 & -0.40 & \textgreater 8.00 & \textgreater 8.00 \\
851180801 & NGC5907 ULX-1 & 4XMMJ151558.6+561810 &  & 4.96 & -0.27 & \textgreater 8.00 & \textgreater 8.00 \\
884220201 & NGC5907 ULX-1 & 4XMMJ151558.6+561810 & \checkmark & 6.69 & 0.81 & \textgreater 8.00 & \textgreater 8.00 \\
884220301 & NGC5907 ULX-1 & 4XMMJ151558.6+561810 &  & 4.72 & 0.46 & \textgreater 8.00 & 6.71 \\
884220401 & NGC5907 ULX-1 & 4XMMJ151558.6+561810 & \checkmark & 5.82 & 0.49 & \textgreater 8.00 & \textgreater 8.00 \\
891801501 & NGC5907 ULX-1 & 4XMMJ151558.6+561810 & \checkmark & 4.41 & -0.21 & \textgreater 8.00 & \textgreater 8.00 \\
893810301 & NGC5907 ULX-1 & 4XMMJ151558.6+561810 &  & 4.27 & 0.63 & \textgreater 8.00 & 6.96 \\* \midrule
693760101 & NGC7793 P13 & 4XMMJ235751.0-323726 &  & -3.89 & -4.72 & \textgreater 8.00 & \textgreater 8.00 \\
693760401 & NGC7793 P13 & 4XMMJ235751.0-323726 & \checkmark & 7.55 & 1.61 & \textgreater 8.00 & \textgreater 8.00 \\
748390901 & NGC7793 P13 & 4XMMJ235751.0-323726 & \checkmark & \textgreater 8.00 & 0.12 & \textgreater 8.00 & \textgreater 8.00 \\
781800101 & NGC7793 P13 & 4XMMJ235751.0-323726 & \checkmark & \textgreater 8.00 & -0.04 & \textgreater 8.00 & \textgreater 8.00 \\
804670201 & NGC7793 P13 & 4XMMJ235751.0-323726 &  & \textgreater 8.00 & 0.86 & \textgreater 8.00 & \textgreater 8.00 \\
804670301 & NGC7793 P13 & 4XMMJ235751.0-323726 & \checkmark & \textgreater 8.00 & 0.44 & \textgreater 8.00 & \textgreater 8.00 \\
804670401 & NGC7793 P13 & 4XMMJ235751.0-323726 & \checkmark & \textgreater 8.00 & \textgreater 8.00 & \textgreater 8.00 & \textgreater 8.00 \\
804670501 & NGC7793 P13 & 4XMMJ235751.0-323726 & \checkmark & \textgreater 8.00 & 0.16 & \textgreater 8.00 & \textgreater 8.00 \\
804670601 & NGC7793 P13 & 4XMMJ235751.0-323726 & \checkmark & \textgreater 8.00 & \textgreater 8.00 & \textgreater 8.00 & \textgreater 8.00 \\
804670701 & NGC7793 P13 & 4XMMJ235751.0-323726 & \checkmark & \textgreater 8.00 & 0.50 & \textgreater 8.00 & \textgreater 8.00 \\
823410301 & NGC7793 P13 & 4XMMJ235751.0-323726 & \checkmark & \textgreater 8.00 & \textgreater 8.00 & \textgreater 8.00 & \textgreater 8.00 \\
823410401 & NGC7793 P13 & 4XMMJ235751.0-323726 & \checkmark & \textgreater 8.00 & 0.51 & \textgreater 8.00 & \textgreater 8.00 \\
840990101 & NGC7793 P13 & 4XMMJ235751.0-323726 & \checkmark & \textgreater 8.00 & 0.61 & \textgreater 8.00 & \textgreater 8.00 \\
853981001 & NGC7793 P13 & 4XMMJ235751.0-323726 & \checkmark & 6.55 & 0.80 & \textgreater 8.00 & \textgreater 8.00 \\
861600101 & NGC7793 P13 & 4XMMJ235751.0-323726 &  & 7.10 & \textgreater 8.00 & \textgreater 8.00 & \textgreater 8.00 \\
883780101 & NGC7793 P13 & 4XMMJ235751.0-323726 &  & -0.62 & -0.98 & \textgreater 8.00 & \textgreater 8.00 \\
883780201 & NGC7793 P13 & 4XMMJ235751.0-323726 &  & -2.58 & -2.17 & \textgreater 8.00 & \textgreater 8.00 \\* \bottomrule
\end{longtable}
\tablefoot{The \checkmark{} symbol indicates that a pulsation was observed in that observation.}

\section{New candidate \acp{PULX}}

\begin{longtable}[c]{@{}cccccc@{}}
\caption{The predicted $z$-scores for the 85 candidate \acp{PULX}, sorted by coordinates.}
\label{tab:complete-table}\\
\toprule
\multirow{2}{*}{{Galaxy}} & \multirow{2}{*}{{4XMMName}} & \multicolumn{4}{c}{{z-score}} \\* \cmidrule(l){3-6} 
 &  & None & Only \lpeak & Only \fpeak & Both \\* \midrule
\endfirsthead
\multicolumn{6}{c}%
{{\bfseries Table \thetable\ continued from previous page}} \\
\toprule
\multirow{2}{*}{{Galaxy}} & \multirow{2}{*}{{4XMMName}} & \multicolumn{4}{c}{{z-score}} \\* \cmidrule(l){3-6} 
 &  & None & Only \lpeak & Only \fpeak & Both \\* \midrule
\endhead
\bottomrule
\endfoot
\endlastfoot
NGC0224 & 4XMMJ004253.1+411423 & 2.69 & -8.13 & > 8.00 & > 8.00 \\
NGC0247 & 4XMMJ004703.8-204743 & 7.68 & -4.22 & > 8.00 & > 8.00 \\
NGC0253 & 4XMMJ004709.1-252123 & 5.88 & 2.00 & > 8.00 & > 8.00 \\
NGC0253 & 4XMMJ004717.5-251812 & > 8.00 & > 8.00 & > 8.00 & > 8.00 \\
NGC0253 & 4XMMJ004722.6-252050 & > 8.00 & > 8.00 & > 8.00 & > 8.00 \\
NGC0253 & 4XMMJ004732.9-251749 & 7.05 & 1.99 & > 8.00 & > 8.00 \\
NGC0470 & 4XMMJ011942.7+032422 & 7.30 & 2.36 & > 8.00 & -8.13 \\
NGC0628 & 4XMMJ013651.1+154546 & 7.32 & > 8.00 & > 8.00 & > 8.00 \\
NGC0891 & 4XMMJ022231.4+422024 & 6.01 & 1.85 & 2.00 & 2.07 \\
NGC0891 & 4XMMJ022231.5+422023 & 6.36 & 1.59 & > 8.00 & > 8.00 \\
NGC0891 & 4XMMJ022233.4+422027 & 5.85 & 2.47 & > 8.00 & > 8.00 \\
NGC0925 & 4XMMJ022727.5+333443 & > 8.00 & 2.25 & > 8.00 & > 8.00 \\
NGC1042 & 4XMMJ024025.6-082429 & 6.97 & 1.67 & > 8.00 & > 8.00 \\
NGC1313 & 4XMMJ031819.9-662910 & > 8.00 & 2.83 & > 8.00 & > 8.00 \\
NGC1313 & 4XMMJ031820.9-663034 & 6.36 & 2.91 & 7.61 & > 8.00 \\
NGC1365 & 4XMMJ033345.4-361014 & 5.60 & > 8.00 & > 8.00 & > 8.00 \\
NGC1399 & 4XMMJ033831.8-352603 & 7.52 & > 8.00 & > 8.00 & > 8.00 \\
NGC1427A & 4XMMJ034012.2-353741 & 5.31 & -4.53 & 4.58 & 4.96 \\
IC0342 & 4XMMJ034555.5+680455 & > 8.00 & 1.18 & > 8.00 & > 8.00 \\
IC0342 & 4XMMJ034615.8+681113 & 7.24 & -2.80 & > 8.00 & > 8.00 \\
NGC1559 & 4XMMJ041730.7-624722 & 3.19 & -3.31 & > 8.00 & > 8.00 \\
NGC2276 & 4XMMJ072647.8+854550 & 4.57 & -1.71 & 5.52 & -8.13 \\
NGC2403 & 4XMMJ073625.5+653539 & 7.68 & 3.07 & > 8.00 & > 8.00 \\
UGC04305 & 4XMMJ081929.0+704219 & > 8.00 & 2.12 & > 8.00 & > 8.00 \\
PGC026378 & 4XMMJ091948.8-121429 & 4.77 & -4.60 & > 8.00 & > 8.00 \\
NGC3031 & 4XMMJ095510.3+690501 & 7.27 & > 8.00 & > 8.00 & > 8.00 \\
NGC3031 & 4XMMJ095524.2+690957 & 4.78 & 0.12 & > 8.00 & > 8.00 \\
NGC3031 & 4XMMJ095532.9+690033 & > 8.00 & 1.50 & > 8.00 & > 8.00 \\
NGC3310 & 4XMMJ103844.9+533004 & 5.32 & 2.30 & 2.26 & 3.24 \\
NGC3621 & 4XMMJ111816.0-324910 & > 8.00 & > 8.00 & > 8.00 & > 8.00 \\
NGC3628 & 4XMMJ112015.7+133514 & 5.23 & -2.43 & > 8.00 & > 8.00 \\
NGC3627 & 4XMMJ112020.8+125847 & 7.19 & 2.58 & > 8.00 & > 8.00 \\
NGC3631 & 4XMMJ112054.3+531040 & 6.28 & 1.40 & 4.71 & 2.50 \\
NGC4038 & 4XMMJ120151.2-185224 & 4.04 & 0.79 & > 8.00 & > 8.00 \\
NGC4038 & 4XMMJ120152.0-185132 & 3.82 & -0.60 & 5.73 & 6.15 \\
NGC4038 & 4XMMJ120154.0-185202 & 6.18 & -1.11 & > 8.00 & > 8.00 \\
NGC4038 & 4XMMJ120156.3-185158 & 3.98 & 1.73 & -0.85 & 1.14 \\
NGC4157 & 4XMMJ121059.6+502831 & 5.94 & > 8.00 & > 8.00 & > 8.00 \\
NGC4406 & 4XMMJ122609.3+125557 & 4.01 & -8.13 & > 8.00 & > 8.00 \\
NGC4449 & 4XMMJ122817.7+440633 & > 8.00 & > 8.00 & > 8.00 & > 8.00 \\
NGC4472 & 4XMMJ122953.4+075936 & 4.49 & -7.29 & 4.21 & 4.16 \\
NGC4490 & 4XMMJ123030.8+413913 & 5.27 & -1.72 & > 8.00 & > 8.00 \\
NGC4490 & 4XMMJ123031.9+413917 & 5.21 & -4.29 & > 8.00 & > 8.00 \\
NGC4490 & 4XMMJ123043.1+413818 & 4.42 & 0.81 & -1.85 & 0.72 \\
NGC4437 & 4XMMJ123242.7+000655 & 5.35 & 0.51 & > 8.00 & > 8.00 \\
NGC4559 & 4XMMJ123551.7+275604 & 7.75 & 2.53 & > 8.00 & > 8.00 \\
NGC4559 & 4XMMJ123558.4+275742 & 7.54 & 2.77 & > 8.00 & > 8.00 \\
NGC4631 & 4XMMJ124155.6+323216 & 6.21 & 2.51 & 4.16 & 4.51 \\
NGC4631 & 4XMMJ124157.4+323203 & 7.19 & > 8.00 & > 8.00 & > 8.00 \\
NGC4649 & 4XMMJ124338.3+113525 & 6.48 & > 8.00 & > 8.00 & > 8.00 \\
NGC4736 & 4XMMJ125048.6+410742 & > 8.00 & > 8.00 & > 8.00 & > 8.00 \\
NGC4939 & 4XMMJ130413.9-102102 & 4.57 & -5.34 & -1.58 & 0.63 \\
NGC4945 & 4XMMJ130538.0-492544 & 5.44 & 1.44 & 6.44 & 6.97 \\
NGC5055 & 4XMMJ131519.5+420301 & > 8.00 & 2.56 & > 8.00 & > 8.00 \\
NGC5055 & 4XMMJ131549.4+420128 & 0.16 & -5.82 & > 8.00 & > 8.00 \\
NGC5128 & 4XMMJ132542.2-425943 & > 8.00 & 2.66 & > 8.00 & > 8.00 \\
NGC5204 & 4XMMJ132938.6+582505 & > 8.00 & 2.71 & > 8.00 & > 8.00 \\
NGC5194 & 4XMMJ132939.7+471239 & > 8.00 & > 8.00 & > 8.00 & > 8.00 \\
NGC5194 & 4XMMJ132943.2+471134 & 4.98 & -5.97 & > 8.00 & > 8.00 \\
NGC5194 & 4XMMJ132953.2+471042 & 6.16 & > 8.00 & > 8.00 & > 8.00 \\
NGC5194 & 4XMMJ133007.5+471106 & 6.77 & 2.72 & 6.86 & 7.28 \\
NGC5236 & 4XMMJ133659.4-294958 & 4.90 & 1.58 & > 8.00 & 7.15 \\
NGC5236 & 4XMMJ133705.1-295206 & 7.69 & 2.42 & > 8.00 & > 8.00 \\
NGC5236 & 4XMMJ133719.8-295348 & 7.70 & 2.75 & > 8.00 & > 8.00 \\
NGC5457 & 4XMMJ140303.8+542735 & > 8.00 & > 8.00 & > 8.00 & > 8.00 \\
NGC5457 & 4XMMJ140314.2+541806 & > 8.00 & > 8.00 & > 8.00 & > 8.00 \\
NGC5408 & 4XMMJ140319.6-412258 & > 8.00 & > 8.00 & > 8.00 & > 8.00 \\
NGC5457 & 4XMMJ140414.1+542605 & 7.43 & -1.94 & > 8.00 & > 8.00 \\
NGC5643 & 4XMMJ143242.2-440939 & 7.65 & 2.02 & > 8.00 & > 8.00 \\
NGC5813 & 4XMMJ150116.5+014134 & 2.37 & -1.94 & > 8.00 & > 8.00 \\
ESO137-034 & 4XMMJ163516.6-580520 & 6.78 & > 8.00 & > 8.00 & > 8.00 \\
ESO101-004 & 4XMMJ163818.7-642201 & 2.62 & -8.13 & 6.44 & -8.13 \\
NGC6215 & 4XMMJ165110.7-585956 & 2.95 & -6.84 & -0.26 & 1.43 \\
NGC6221 & 4XMMJ165251.5-591503 & -2.20 & -7.59 & > 8.00 & > 8.00 \\
NGC6643 & 4XMMJ181955.6+743446 & 6.19 & -2.45 & > 8.00 & > 8.00 \\
NGC6643 & 4XMMJ181959.2+743513 & -0.21 & -4.47 & 3.84 & 3.64 \\
NGC6702 & 4XMMJ184700.0+454226 & 1.44 & -4.24 & > 8.00 & -8.13 \\
NGC6946 & 4XMMJ203456.8+600811 & 7.49 & > 8.00 & > 8.00 & > 8.00 \\
NGC6946 & 4XMMJ203500.1+600908 & > 8.00 & > 8.00 & > 8.00 & > 8.00 \\
NGC6946 & 4XMMJ203500.7+601130 & > 8.00 & > 8.00 & > 8.00 & > 8.00 \\
IC5052 & 4XMMJ205216.9-691316 & 6.42 & 2.26 & > 8.00 & > 8.00 \\
NGC7090 & 4XMMJ213631.8-543357 & 5.18 & -0.56 & 5.32 & 5.18 \\
NGC7090 & 4XMMJ213633.8-543400 & 2.69 & -3.25 & 4.24 & 4.01 \\
NGC7793 & 4XMMJ235747.9-323457 & > 8.00 & 3.01 & > 8.00 & > 8.00 \\
NGC7793 & 4XMMJ235808.8-323403 & 5.94 & 2.68 & > 8.00 & > 8.00 \\* \bottomrule
\end{longtable}
\tablefoot{For each candidate \ac{PULX}, we report only the observation associated with the highest $z$-score when considering only \fpeak{}.}

 \section{Subdivision of candidate \acp{PULX} in clusters}

 \begin{longtable}[c]{@{}cccc@{}}
 \caption{Number of observations classified by the clustering algorithm as candidates (cluster $C_P$) and non-candidates (cluster $C_U$) for each source (excluding known \acp{PULX}) and the percentage of observations identified as candidates relative to the total number of observations per source. }
 \label{tab:distribution-candidates}\\
 \toprule
 4XMMName & Candidates & Not candidates & \% candidates \\* \midrule
 \endfirsthead
 \multicolumn{4}{c}%
 {{\bfseries Table \thetable\ continued from previous page}} \\
 \toprule
 4XMMName & Candidates & Not candidates & \% candidates \\* \midrule
 \endhead
 \bottomrule
 \endfoot
 \endlastfoot
4XMMJ031819.9-662910 & 32 & 0 & 100\% \\
4XMMJ095532.9+690033 & 22 & 0 & 100\% \\
4XMMJ004253.1+411423 & 20 & 0 & 100\% \\
4XMMJ081929.0+704219 & 18 & 0 & 100\% \\
4XMMJ095524.2+690957 & 17 & 0 & 100\% \\
4XMMJ203500.7+601130 & 16 & 0 & 100\% \\
4XMMJ235808.8-323403 & 16 & 0 & 100\% \\
4XMMJ140319.6-412258 & 13 & 0 & 100\% \\
4XMMJ004703.8-204743 & 11 & 0 & 100\% \\
4XMMJ004732.9-251749 & 9 & 0 & 100\% \\
4XMMJ073625.5+653539 & 9 & 0 & 100\% \\
4XMMJ132938.6+582505 & 9 & 0 & 100\% \\
4XMMJ024025.6-082429 & 7 & 0 & 100\% \\
4XMMJ022231.5+422023 & 6 & 0 & 100\% \\
4XMMJ022233.4+422027 & 6 & 0 & 100\% \\
4XMMJ034555.5+680455 & 6 & 0 & 100\% \\
4XMMJ034615.8+681113 & 6 & 0 & 100\% \\
4XMMJ122817.7+440633 & 3 & 0 & 100\% \\
4XMMJ123030.8+413913 & 3 & 0 & 100\% \\
4XMMJ131519.5+420301 & 3 & 0 & 100\% \\
4XMMJ150116.5+014134 & 3 & 0 & 100\% \\
4XMMJ165251.5-591503 & 3 & 0 & 100\% \\
4XMMJ022727.5+333443 & 2 & 0 & 100\% \\
4XMMJ123031.9+413917 & 2 & 0 & 100\% \\
4XMMJ123551.7+275604 & 2 & 0 & 100\% \\
4XMMJ123558.4+275742 & 2 & 0 & 100\% \\
4XMMJ130538.0-492544 & 2 & 0 & 100\% \\
4XMMJ132542.2-425943 & 2 & 0 & 100\% \\
4XMMJ140314.2+541806 & 2 & 0 & 100\% \\
4XMMJ163818.7-642201 & 2 & 0 & 100\% \\
4XMMJ034012.2-353741 & 1 & 0 & 100\% \\
4XMMJ041730.7-624722 & 1 & 0 & 100\% \\
4XMMJ072647.8+854550 & 1 & 0 & 100\% \\
4XMMJ091948.8-121429 & 1 & 0 & 100\% \\
4XMMJ111816.0-324910 & 1 & 0 & 100\% \\
4XMMJ112015.7+133514 & 1 & 0 & 100\% \\
4XMMJ112020.8+125847 & 1 & 0 & 100\% \\
4XMMJ121059.6+502831 & 1 & 0 & 100\% \\
4XMMJ122953.4+075936 & 1 & 0 & 100\% \\
4XMMJ123242.7+000655 & 1 & 0 & 100\% \\
4XMMJ124155.6+323216 & 1 & 0 & 100\% \\
4XMMJ124157.4+323203 & 1 & 0 & 100\% \\
4XMMJ125048.6+410742 & 1 & 0 & 100\% \\
4XMMJ130413.9-102102 & 1 & 0 & 100\% \\
4XMMJ131549.4+420128 & 1 & 0 & 100\% \\
4XMMJ163516.6-580520 & 1 & 0 & 100\% \\
4XMMJ165110.7-585956 & 1 & 0 & 100\% \\
4XMMJ181959.2+743513 & 1 & 0 & 100\% \\
4XMMJ203456.8+600811 & 1 & 0 & 100\% \\
4XMMJ205216.9-691316 & 1 & 0 & 100\% \\
4XMMJ213631.8-543357 & 1 & 0 & 100\% \\
4XMMJ235747.9-323457 & 1 & 0 & 100\% \\
4XMMJ004717.5-251812 & 6 & 1 & 86\% \\
4XMMJ132943.2+471134 & 7 & 2 & 78\% \\
4XMMJ140303.8+542735 & 3 & 1 & 75\% \\
4XMMJ133705.1-295206 & 4 & 2 & 67\% \\
4XMMJ122609.3+125557 & 2 & 1 & 67\% \\
4XMMJ143242.2-440939 & 2 & 1 & 67\% \\
4XMMJ123043.1+413818 & 3 & 2 & 60\% \\
4XMMJ132953.2+471042 & 5 & 4 & 56\% \\
4XMMJ133719.8-295348 & 5 & 4 & 56\% \\
4XMMJ004709.1-252123 & 4 & 4 & 50\% \\
4XMMJ140414.1+542605 & 2 & 2 & 50\% \\
4XMMJ011942.7+032422 & 1 & 1 & 50\% \\
4XMMJ124338.3+113525 & 1 & 1 & 50\% \\
4XMMJ004722.6-252050 & 4 & 5 & 44\% \\
4XMMJ120154.0-185202 & 3 & 4 & 43\% \\
4XMMJ095510.3+690501 & 5 & 8 & 38\% \\
4XMMJ120152.0-185132 & 3 & 5 & 38\% \\
4XMMJ013651.1+154546 & 1 & 2 & 33\% \\
4XMMJ033345.4-361014 & 1 & 2 & 33\% \\
4XMMJ033831.8-352603 & 1 & 2 & 33\% \\
4XMMJ103844.9+533004 & 1 & 2 & 33\% \\
4XMMJ184700.0+454226 & 1 & 2 & 33\% \\
4XMMJ213633.8-543400 & 1 & 2 & 33\% \\
4XMMJ132939.7+471239 & 3 & 8 & 27\% \\
4XMMJ112054.3+531040 & 1 & 3 & 25\% \\
4XMMJ022231.4+422024 & 1 & 4 & 20\% \\
4XMMJ031820.9-663034 & 1 & 4 & 20\% \\
4XMMJ181955.6+743446 & 1 & 4 & 20\% \\
4XMMJ120156.3-185158 & 1 & 6 & 14\% \\
4XMMJ203500.1+600908 & 2 & 13 & 13\% \\
4XMMJ120151.2-185224 & 1 & 7 & 13\% \\
4XMMJ133659.4-294958 & 1 & 8 & 11\% \\
4XMMJ133007.5+471106 & 1 & 10 & 9\% \\* \bottomrule
 \end{longtable}

\end{appendix}

\end{document}